\documentclass[]{elsarticle}

\PassOptionsToPackage{table}{xcolor}

\renewcommand*{\cite}{\citep}

\usepackage{tikz}
\usepackage{etoolbox}
\usepackage{xfp}
\usepackage{xcolor}
\usepackage{amsmath}
\usepackage[many]{tcolorbox}
\usepackage{relsize}
\usepackage{hyperref}
\usepackage{cleveref}
\usepackage{balance}
\usepackage{import}
\usepackage{booktabs}
\usepackage{longtable}
\usepackage{lscape}
\usepackage{multicol}
\usepackage[framemethod=TikZ]{mdframed}
\usepackage{dashrule}
\usepackage{fontawesome5}
\usepackage{moresize}

\usepackage{diagbox} 
\usepackage{graphicx}
\usepackage{caption}
\graphicspath{{./images/}}

\usepackage{tikz}

\newcommand*\circled[1]{\tikz[baseline=(char.base)]{
\node[minimum size=\baselineskip,shape=circle,draw,inner sep=1pt,font=\footnotesize,fill=tablebg,] (char) {#1};}}

\usepackage[shortcuts]{extdash}
\usepackage{xifthen} 
\newcommand{\citedata}[2][]{%
\ifthenelse{\isempty{#1}}{\href{http://stackoverflow.com/questions/#2}{\textit{#2}}}{%
\ifthenelse{\isempty{#2}}%
{\unskip~``\textsc{#1}"}%
{\unskip~``\textsc{#1}"\allowbreak{}(\href{http://stackoverflow.com/questions/#2}{\textit{#2}})}}%
}

\newcommand{\stackoverflow}{Stack Overflow}

\newcommand{\descriptionrow}[1]{\multicolumn{2}{p{\textwidth}}{#1}}

\begin{document}
\let\WriteBookmarks\relax
\def\floatpagepagefraction{1}
\def\textpagefraction{.001}


\title{Sentiment Overflow in the Testing Stack: \\
Analysing Software Testing Posts on Stack Overflow}

\pagenumbering{roman}

\definecolor{pastelgreen}{HTML}{ccebc5}
\definecolor{pastelyellow}{HTML}{fed9a6}
\definecolor{pastelred}{HTML}{fbb4ae}
\definecolor{pastelrose}{HTML}{fddaec}
\definecolor{pastelgray}{HTML}{f2f2f2}
\definecolor{tablebg}{HTML}{f0f0f0}

\tikzstyle positivebar=[fill=pastelgreen,draw=black]
\tikzstyle bothbar=[fill=pastelyellow,draw=black]
\tikzstyle negativebar=[fill=pastelred,draw=black!75]
\tikzstyle neutralbar=[fill=pastelgray,draw=black]
\tikzstyle scalex=[very thin,draw=black!75]
\tikzstyle scalex2=[draw=black!75]

\newcommand{\fourstackedbarslegend}[1][5]{
\begin{tikzpicture}%
   \draw[positivebar] (0cm,0cm) rectangle (
       \fpeval{#1 / 4 } , 0.335);
   \draw[scalex2](0.025,0.075) node[text width=0pt, text height=0pt, font=\footnotesize] {\tiny Positive};

   \draw[bothbar] (\fpeval{#1/4}cm,0cm) rectangle (
       \fpeval{ \fpeval{#1 / 4 } + \fpeval{ #1 / 4 } -.15 },
       0.335);
   \draw[scalex2](
       \fpeval{ \fpeval{#1 / 4 } + 0.025},
          0.075) node[text width=0pt, text height=0pt, font=\footnotesize] {\tiny Both};

      \draw[negativebar] (\fpeval{(#1 /4)*2}-.15,0cm) rectangle (
       \fpeval{ \fpeval{#1 / 4 } * 2 + \fpeval{ #1 / 4 } },
       0.335);
   \draw[scalex2](
       \fpeval{ \fpeval{#1 / 4 } * 2 + 0.025 -.15 },
          0.075) node[text width=0pt, text height=0pt, font=\footnotesize] {\tiny Negative};

  \draw[neutralbar] ( \fpeval{ (#1 / 4) * 3 }, 0cm ) rectangle ( #1, 0.335);
   \draw[scalex2](
       \fpeval{ \fpeval{#1 / 4 } * 3 + 0.025 },
          0.075) node[text width=0pt, text height=0pt, font=\footnotesize] {\tiny Neutral};
      \draw[scalex2](\fpeval{#1 + 0.1}, 0.075) node[text width=0pt, text height=0pt, font=\footnotesize] {($\Sigma$)};

   \end{tikzpicture}%
}
\newcommand{\fourstackedbars}[5][5]{%
\begin{tikzpicture}%
   \draw[positivebar] (0cm,0cm) rectangle (
       \fpeval{ \fpeval{#1 / \fpeval{#2+#3+#4+#5}} * #2 } ,
       0.335);
   \ifthenelse{\equal{#2}{0}}{}{
   \draw[scalex2](
    \fpeval{
     \fpeval{ \fpeval{#1 / \fpeval{#2+#3+#4+#5}} * #2 }
        / 2 - 0.1},
          0.075) node[text width=0pt, text height=0pt, font=\footnotesize] {#2};
      }

   \draw[bothbar] (
       \fpeval{ \fpeval{#1 / \fpeval{#2+#3+#4+#5}} * #2 } ,
       0cm ) rectangle (
       \fpeval{
           \fpeval{ \fpeval{#1 / \fpeval{#2+#3+#4+#5}} * #2 }
           + \fpeval{ \fpeval{#1 / \fpeval{#2+#3+#4+#5}} * #3 }
       }, 0.335);
   \ifthenelse{\equal{#3}{0}}{}{
   \draw[scalex2](
        \fpeval{
                \fpeval{ \fpeval{#1 / \fpeval{#2+#3+#4+#5}} * #3}
                / 2
                - 0.1
                + \fpeval{
                      \fpeval{#1 / \fpeval{#2+#3+#4+#5}} * #2}
                },
          0.075) node[text width=0pt, text height=0pt, font=\footnotesize] {#3};
  }
   \draw[negativebar] (
       \fpeval{
           \fpeval{ \fpeval{#1 / \fpeval{#2+#3+#4+#5}} * #2 }
           + \fpeval{ \fpeval{#1 / \fpeval{#2+#3+#4+#5}} * #3 }
       }, 0cm ) rectangle (
       \fpeval{
           \fpeval{ \fpeval{#1 / \fpeval{#2+#3+#4+#5}} * #2 }
           + \fpeval{ \fpeval{#1 / \fpeval{#2+#3+#4+#5}} * #3 }
           + \fpeval{ \fpeval{#1 / \fpeval{#2+#3+#4+#5}} * #4 }
       }, 0.335);
   \ifthenelse{\equal{#4}{0}}{}{
   \draw[scalex2](
        \fpeval{
                \fpeval{ \fpeval{#1 / \fpeval{#2+#3+#4+#5}} * #4}
                / 2
                - 0.1
                + \fpeval{
                      \fpeval{#1 / \fpeval{#2+#3+#4+#5}} * #2
                      + \fpeval{#1 / \fpeval{#2+#3+#4+#5}} * #3
                  }
                },
          0.075) node[text width=0pt, text height=0pt, font=\footnotesize] {#4};
  }

   \draw[neutralbar] (
       \fpeval{
           \fpeval{ \fpeval{#1 / \fpeval{#2+#3+#4+#5}} * #2 }
           + \fpeval{ \fpeval{#1 / \fpeval{#2+#3+#4+#5}} * #3 }
           + \fpeval{ \fpeval{#1 / \fpeval{#2+#3+#4+#5}} * #4 }
       }, 0cm ) rectangle (#1, 0.335);
   \ifthenelse{\equal{#5}{0}}{}{
   \draw[scalex2](
        \fpeval{
                \fpeval{ \fpeval{#1 / \fpeval{#2+#3+#4+#5}} * #5}
                / 2
                - 0.1
                + \fpeval{
                      \fpeval{#1 / \fpeval{#2+#3+#4+#5}} * #2
                      + \fpeval{#1 / \fpeval{#2+#3+#4+#5}} * #3
                      + \fpeval{#1 / \fpeval{#2+#3+#4+#5}} * #4
                  }
                },
          0.075) node[text width=0pt, text height=0pt, font=\footnotesize] {#5};
  }
  \draw[scalex2](\fpeval{#1 + 0.1}, 0.075) node[text width=0pt, text height=0pt, font=\footnotesize] {(\fpeval{#2+#3+#4+#5})};
   \end{tikzpicture}%
}

\newcounter{fcode}
\refstepcounter{fcode}
\makeatletter
\newcommand{\fcode}[2]{%
    \phantomsection
    #1 (F.#2)\def\@currentlabel{\unexpanded{(F.#2)}}\label{fcode:#2}%
}
\newcommand{\category}[2]{%
    \phantomsection
    #1 (C.#2)\def\@currentlabel{\unexpanded{(C.#2)}}\label{cat:#2}%
}
\makeatother

\newenvironment{textbox}[2][]{%
\ifstrempty{#1}
{\mdfsetup{innerleftmargin=10pt, innerrightmargin=10pt,%
frametitle={%
\tikz[baseline=(current bounding box.east), outer sep=0pt]
\node[anchor=east,rectangle,fill=tablebg]
{\strut Key insight};}}
}%
{\mdfsetup{innerleftmargin=10pt, innerrightmargin=10pt,%
frametitle={%
\tikz[baseline=(current bounding box.east), outer sep=0pt]
\node[anchor=east,rectangle,draw=black!75,fill=tablebg]
{\textcolor{black}{~#1~}};}}%
}%
\mdfsetup{innertopmargin=4pt,innerbottommargin=5pt, linecolor=black!75,%
linewidth=.5pt,topline=true,roundcorner=5pt,skipabove=5pt,%
frametitleaboveskip=\dimexpr-\ht\strutbox\relax
}
\begin{mdframed}[]\relax%
\label{#2}}{\end{mdframed}}

\NewDocumentCommand \ColoredBadge { m m m m }
  {%
  \ifstrempty{#4}
  {%
    \tcbsidebyside
      [
        enhanced, sidebyside, sidebyside adapt=both,
        bicolor, colback=black, colbacklower=green!70!black,
        frame hidden,
        left=2pt, right=2pt, top=0pt, bottom=0pt,
        sidebyside gap=1em,
      ]
    {\textcolor{gray}{#1}~\textcolor{white}{#2}\hspace{-.5em}\phantom{q}}
    {\sffamily\hspace{-1em}\phantom{#1}\textcolor{white}{#3}\hspace{-.5em}\phantom{q}}
  }
  {
    \tcbsidebyside
      [
        enhanced, sidebyside, sidebyside adapt=both,
        bicolor, colback=black, colbacklower={#4},
        frame hidden,
        left=2pt, right=2pt, top=0pt, bottom=0pt,
        sidebyside gap=1em,
      ]
    {\textcolor{gray}{#1}~\textcolor{white}{#2}\hspace{-.5em}\phantom{q}}
    {\sffamily\hspace{-1.25em}\phantom{#1}\textcolor{white}{#3}\hspace{-.5em}\phantom{q}}
  }
  }


\makeatletter

\newcommand{\review}[4]{%
\phantomsection%
\noindent\ifthenelse{\equal{#3}{}}{\faIcon[regular]{square}}{\faIcon{pen-square}}
\ifthenelse{\equal{#4}{}}{}{}
    R.#2: #1%
    \def\@currentlabel{\unexpanded{(R.#2)}}%
    \label{review:#2}%
}
\newcommand\revised[2]{{#1}}
\makeatother


\makeatletter

\newcommand{\reviewtwo}[4]{%
\phantomsection%
\noindent\ifthenelse{\equal{#3}{}}{\faIcon[regular]{square}}{\faIcon{pen-square}}
\ifthenelse{\equal{#4}{}}{}{}
    R2.#2: #1%
    \def\@currentlabel{\unexpanded{(R2.#2)}}%
    \label{review2:#2}%
}
\newcommand\revisedtwo[2]{{#1}}
\makeatother

\makeatletter
\newcommand{\csvdel}{,}
\newcommand{\reflist}[1][,]{
    \renewcommand{\csvdel}{\renewcommand{\csvdel}{{\tiny#1\,\allowbreak}}}
\csname decl@#1\endcsname{\tiny[}\checknextarg}%
\newcommand{\checknextarg}{\@ifnextchar\bgroup{\gobblenext}{}}
\newcommand{\gobblenext}[1]{\csvdel{{\tiny\citedata[]{#1}}}\@ifnextchar\bgroup{\gobblenext}{{\tiny]}}}
\makeatother



\author[1]{Mark Swillus\fnref{fn1}}
\ead{m.swillus@tudelft.nl}
\fntext[fn1]{~\faIcon{orcid} \href{https://orcid.org/0000-0003-3746-1030}{0000-0003-3746-1030}}

\address[1]{Delft University of Technology, Van Mourik Broekmanweg 6, 2628XE Delft, The~Netherlands}

\author[1]{Andy Zaidman\fnref{fn2}}
\fntext[fn2]{~\faIcon{orcid} \href{https://orcid.org/0000-0003-2413-3935}{0000-0003-2413-3935}}
\ead{a.e.zaidman@tudelft.nl}

\begin{abstract}
Software testing is an integral part
of modern software engineering practice.
Past research has not only underlined its significance,
but also revealed its multi-faceted nature.
The practice of software testing
and its adoption
is influenced by many factors
that go beyond tools or technology.
This paper sets out
to investigate the context
of software testing
from the practitioners' point of view
by mining and analyzing sentimental posts
on the widely used
question and answer website \stackoverflow{}.
By qualitatively analyzing
sentimental expressions of practitioners,
which we extract from the \stackoverflow{} dataset
using sentiment analysis tools,
we discern factors
that help us to better understand
the lived experience
of software engineers
with regards to software testing.
Grounded in the data that we have analyzed,
we argue
that sentiments like
insecurity, despair and aspiration,
have an impact
on practitioners' attitude
towards testing.
We suggest that they are connected
to concrete factors
like the level of complexity of projects
in which software testing is practiced.
\end{abstract}


\maketitle
\pagenumbering{arabic}
\setcounter{page}{1}

\section{Introduction}
We already know for over 40 years that software testing is one of the most pragmatic mechanisms
by which we can ensure the quality of the software artefacts that
we engineer~\cite{Athanasiou_test_2014,carstensen_lets_1995,hetzel_complete_1988,myers_art_2012,yourdon_managing_1988,zaidman_mining_2008,zaidman_studying_2011}.
In the light of the unquestionable growing impact that software and software supported devices are
having on our daily lives, the role of software testing becomes ever more important.
Just consider the year 2017, which has been earmarked
``The Year That Software Bugs Ate the World'' because of the astonishing software failures that
cost the economy \$1.7 trillion in 2017 alone~\cite{mccracken_year_2017}. Crucially, \citet{ko_thirty_2014}
report on software failures that can be directly linked to the loss of 1,500 human lives.
However, to this day there is a schism between widespread recommendations for software engineering practice
and our knowledge of how software testing \emph{actually} happens. The urgency to solve
this conflict was also signalled by others with a call to arms to better understand the testing
process~\cite{bertolino_Software_2007,mantyla_who_2012}.

We have recently seen studies emerge that have observed how software developers test.
\citet{beller_developer_2019,beller_when_2015,beller_how_2015} have investigated when and how developers write test cases in
their Integrated Development Environment.
They observed that around 50\% of the studied projects do not employ automated testing methods
at all. But they also found out that for almost all cases testing happens far less frequently
than developers estimate.
If testing is truly considered a last line of defense against software defects,
we need to understand why developers \emph{do} or \emph{do not}
engineer and execute test cases.

We have already seen glimpses of this in literature. Studies have shown that company culture or
time pressure leads to cognitive biases during testing~\cite{Mohanani_cognitive_2020,Calikli_influence_2013,Salman_what_2022},
estimations of the time it takes to write test are often
inaccurate~\cite{beller_developer_2019, kasurinen_analysis_2009},
availability of documentation shapes the development of tests~\cite{aniche_how_2022}, and that the
cost/benefit of testing is often unclear~\cite{begel_analyze_2014}.
Additionally, \citet{kasurinen_analysis_2009}, \citet{runeson_survey_2006}, and \citet{daka_survey_2014}
highlight issues with motivating developers to test software:
only half of them have positive feelings about testing, and approachability of tools is a
major factor.
Like Prado and Vincenzi~\cite{prado_towards_2018} who studied the perspective
of developers
during the review process of unit tests
to build tools that encourage testing,
we follow
and put the human into the center of attention.
This paper sets out to investigate the circumstances that
influence software engineers
when engineering tests going beyond technical aspects of the
discipline.
\revised{%
Like \citet{sharp_software_2000}, we believe that
in order to improve the discipline,
it is essential to understand
the socio-technical world in which software engineering
is practiced.
For example,
software development practices
which form social circumstances,
like pair programming,
are very likely to have an impact on testing.}{1.1}
To gain a broad overview of what these circumstances are,
we take negative and positive sentiments on the process of automated testing as a proxy.
To gather documents
which describe the experience
of software developers
from their point
of view,
we mine the most popular question and answer platform for
software engineers, namely \textit{\stackoverflow{}}~\cite{baltes_usage_2019}.

\begin{textbox}[RQ1]{}
    How do software engineers express sentiment about testing on \stackoverflow{}?
\end{textbox}

\revised{On the Q\&A platform \stackoverflow{},
on which social interaction plays a key role,
practitioners ask questions about software development
which are answered by a global community of
software developers~\cite{mustafa_what_2022}.}{1.1}
Others have used the \stackoverflow{} dataset
to investigate technical and non-technical aspects
of software engineering.
For example,~\citet{lopez_investigation_2018}
have analyzed security questions on \stackoverflow{},
and provide an overview of the most discussed topics
but also discuss the way in which authors
discuss security questions.
Our goal is to identify factors
that affect practitioners and influence adoption
of, or attitude towards testing.
\revised{We hypothesize that an analysis of sentimental
content on \stackoverflow{}
not only reveals technical factors that can influence adoption or attitudes,
but also descriptions of the social or human context
of practice.
To identify those socio-technical factors,
we deeply examine 200 testing related questions on \stackoverflow{}
instead of analyzing the whole dataset quantitatively.}{1.1}
\revised{We do not only
scrutinize the question
asked by the practitioner,
but also incorporate
answers, comments and the edit history of questions into our analysis.}{3.1}
\revised{%
Going beyond an analysis of
\textit{questions} about technical issues,
we focus on the broader context
that causes sentiment in practitioners.}{1.1}
\revised{%
We therefore
use the term \textit{post}
instead of \textit{question}
to refer to the documents we analyzed
for the remainder of this paper.}{3.1}

\begin{textbox}[RQ2]{}
    Which factors affect sentiment of software engineers towards testing practices?
\end{textbox}
From research done by other authors we know that only a small fraction
of posts on \stackoverflow{} contains strong opinions and emotional statements as they mostly
discuss how to use a piece of technology~\cite[]{lin_sentiment_2018,sengupta_learning_2020}.
This motivates us to create an emotionally rich subset by filtering the dataset using a
semi-automated approach that employs sentiment analysis tools.

\revised{%
To answer both research questions
we apply strategies
of Hoda's basic stage
for socio-technical grounded theory (STGT)~\cite{hoda_socio-technical_2022}
with a constructivist stance as suggested by~\citet{charmaz_constructing_2014}.
STGT provides us with a framework to venture into a broad analysis
of testing practice,
seen not only as a technical phenomenon,
but as a phenomenon in which social factors play an essential role.}{1.4}
\revised{We focus our analysis on the socio-technical
dimension of posts on \stackoverflow{} and show that such an
analysis indeed reveals descriptions of social aspects.
Our analysis informs about issues that contribute
to problems and attitudes towards software testing.}{1.1}
\revised{\revised{More concretely, we analyze the dataset which consists of 200 posts
using initial and focused coding and
techniques for systematic comparison of posts, codes and memos
like diagramming and clustering.
Concluding this paper with a presentation of preliminary
categories and a preliminary interpretive theory,
we motivate consecutive targeted data collection
(theoretic sampling)
to test and extend our analysis and conclusions.}{1.4}}{1.3}
Grounded in the data we analyze,
this paper makes the following contributions:

\begin{itemize}
\item We discuss preliminary hypotheses which explore stimuli and
    inhibitors to testing at a socio-technical level
\item We present a computer aided approach for qualitative analysis of sentimental
    expressions in big datasets
\item \revised{We motivate a research agenda that includes concrete ideas for targeted data
    collection (theoretic sampling) to develop a mature theory of stimuli and inhibitors of software
testing that go beyond tools and technology}{1.3}
\end{itemize}

\section{Background}

\subsection{Sentiment Analysis}

Sentiment analysis is the computational study of opinions, sentiments
and emotions expressed in text.
It essentially tries to infer people's sentiments based on
their language expressions. Sentiment \textit{classification} is a widely studied research
topic of sentiment analysis that focuses on the classification of opinionated
documents as expressing positive or negative opinion~\cite{indurkhya_handbook_2010}.
Automatic classification of sentiment
has been applied in various fields of research over the past
20 years as access to vast amounts of written text about various topics
have become available through the internet.
Already in 1999 \citet{wiebe_development_1999} worked on a dataset for automatic
classification of news articles to identify whether
information is being presented as fact or opinion. While sentiment analysis
is still being used to analyze media platforms like those of news agencies~\cite{balahur_rethinking_2009,pelicon_zero-shot_2020},
its application today also includes
platforms on which a wide variety of people contribute content such as social media or
internet forums. Here sentiment analysis has been used recently to identify personal
attacks or obscene behavior of users~\cite{Saeed_overlapping_2018}.

Techniques for sentiment analysis have also been applied in the context of software
engineering. \citet{mantyla_bootstrapping_2017} analyzed sentiment in comments
on the Jira issue tracker to detect burnout among software developers.
They calculated sentiment scores for each sentence using a dictionary that contains
ratings for the affective meaning of 13,915 English words.
Despite their positive results, they have also raised the issue,
echoed by others~\cite{jongeling_negative_2017}, that general purpose sentiment analysis
tools lack precision when applied to the domain of software engineering.
\citet{lin_sentiment_2018} even question
the validity of all quantitative studies in software engineering
based on sentiment analysis tools as they demonstrate how hard it is to reproduce
results. For example, they judge that there is still a long way to go before
researchers and practitioners can use state-of-the-art sentiment analysis tools to
identify the sentiment expressed in \stackoverflow{} discussions.
\revisedtwo{%
To stimulate more research into the direction of
sentiment analysis, they published the
dataset that was developed in~\cite{lin_sentiment_2018}
which contains 1,500 annotated sentences.
Similarly, to support empirical research in the direction
of emotion detection,
\citet{novielli_gold_2018} developed a dataset
containing 4,800 \stackoverflow{} posts.
Motivated by these voices of criticism and encouragement,
others then tried to develop tools tailored to the domain
of software engineering like Islam et al.,
who have developed the dictionary-based tool DEVA~\cite{islam_deva_2018},
and a machine-learning based tool called MarValous that focuses on emotion
detection~\cite{islam_marvalous_2019}.
In that same period the SentiStrength tool,
which already existed
as a general purpose tool
for sentiment classification,
was tweaked for an application in the
domain of software engineering
by \citet{ahmed_senticr_2017},
who created the tool SentiCR.
In 2020 \citet{zhang_sentiment_2020} address the issue again,
comparing the accuracy of this new
generation of tailor-made sentiment analysis tools
for software engineering
with the accuracy that
deep neural network architectures,
namely transformer models, achieve.
They suggest that transformer models like RoBERTa~\cite{liu_roberta_2019}
are indeed one big step forward
on the long way towards reliable results in sentiment classification
for software engineering.
Finally, in 2021,
\citet{lin_opinion_2022} summarize the knowledge gained
in one decade of research for opinion mining tools.
Among other insights into the field
they provide a guideline for the selection,
usage and evaluation of opinion mining tools for software engineering research.}{1.2}

\subsection{Grounded Theory}%
\label{sub:Grounded Theory}
Grounded theory (GT) is an analytic approach used to
construct ethnographic knowledge~\cite{deener_architecture_2018}.
Its framework is made up of data-gathering techniques and strategies to analyze data.
What distinguishes GT from other approaches is its iterative nature.
While theory development progresses, the GT approach alternates between data collection
and analysis to sustain a high level of involvement with the data~\cite{charmaz_constructing_2014}.

GT was suggested as an approach for qualitative research by
~\citet{glaser_discovery_2010} and has been reinterpreted by different
scholars, resulting in the development of different flavours of GT.
Flavours of GT differ in details on how to execute
techniques and how tightly strategies need to be followed\footnote{For a more elaborate
discussion of the historic development of GT and a complete
comparison of its flavours see Charmaz~\cite[p.~4]{charmaz_constructing_2014} and
Hoda\cite[p.~9]{hoda_socio-technical_2022}}.
Crucially, they also rest on different epistemological stances. Where the original
\textit{Glaserian GT} takes an objective, positivist stance,
\textit{Constructivist GT} proposed by Charmaz, for example,
acknowledges the researchers' subjective perspective.
Constructivist GT moves away from positivism, incorporating the beliefs and preconceptions
of the researcher into analysis.
Situating the GT approach into the field of software engineering research,
Hoda has recently proposed another flavour of GT.
She designed \textit{Socio-technical GT (STGT)} to ease application of GT in her field,
where researchers often struggle to understand and apply it~\cite{hoda_socio-technical_2022}.
With STGT, Hoda proposes to divide GT into distinct phases.
Embracing the iterative nature of GT, STGT encourages exploration in a \textit{Basic Stage}
and helps the socio-technical researcher to transition into an \textit{Advanced Stage}
of theory development.
The separation into those two stages,
which are accompanied by
lean and focused literature reviews,
help the socio-technical researcher
to cover epistemological blind-spots.
All flavours of GT
use comparative- (e.g., \textit{clustering}, \textit{diagramming}), and analytical methods
(e.g., \textit{coding}, \textit{memo writing}), that
are accompanied by a continuous collection of new data samples (\textit{theoretical sampling})
to saturate emerging categories that describe data and to enable the development of
\textit{mature theories} which transparently emerge from the data.
Regarding the analysis of documents, which we set out to do in this paper,
Charmaz states that GT of documents is able to address not only content but also
their audience, production and presentation.
Analysis of documents can reveal what and whom they affect, as they do not only serve as
records but explore, explain, justify and/or foretell
actions~\cite[p.~46]{charmaz_constructing_2014}.

In this paper we follow Hoda's STGT
and present the results of the \textit{Basic Stage}
of our STGT study.
Publication of emerging results
of this exploratory phase is
encouraged by both Hoda and Charmaz.
GT guidelines describe steps
and a path through a long research process.
Depending on
the task and project at hand,
GT invites using those steps flexibly
to raise the analysis to the
desired level of theory construction~\cite{charmaz_constructing_2014}.
Within the framework of STGT
we use strategies and epistemology from
Charmaz Constructivist GT,
raising the data analysis of our dataset
that we take from \stackoverflow{} to a \textit{preliminary theory}.
We present our work
following Hoda's
recommendation who states
that publication even of partial results
is important to receive feedback
from both practitioners and
the research community
to assesses relevance and improve rigour~\cite{hoda_socio-technical_2022}.

\subsection{\stackoverflow{}}%
\label{sub:Researcher Background}
\textit{\stackoverflow{}} is the most popular question- and answer website for software
developers~\cite{baltes_usage_2019}. The website has become an important resource that often
complements official documentation of software libraries and tools.
Its strong presence on search engines,
where a link to the website is very often shown on the first page of results when searching
for software development related topics, indicates its reach that goes far beyond the
17 million registered users~\cite{barzilay_facilitating_2013}.
Studies that often use the official and open \stackoverflow{} dataset,
have underlined the prominence of \stackoverflow{} by showing for example
that 11\% of open source software projects on GitHub that were analyzed in a large scale
field study contain source code snippets that were copied
from \stackoverflow{}~\cite{baltes_usage_2019}.
Over 22 million questions that often contain such code snippets were posted by users
in a wide range of topics that are related to software engineering
since its launch in 2008~\footnote{\href{https://stackexchange.com/sites?view=list\#traffic}{\faIcon{stack-exchange} stackexchange.com/sites}}.
Apart from contributions in the form of questions and answers, users are
also encouraged to take part in moderation efforts.
Up- and down voting, tagging and editing of questions and answers
is rewarded with badges, medals and reputation points.
\revised{Questions on \stackoverflow{} generate living documents,
which are edited by their authors and moderators,
updated, and extended with comments
sometimes even a decade
after they were asked.
Questions and their answers can thus take
the character of knowledge base articles.
\citet{barzilay_facilitating_2013} even argue that the moderation and reward system
has transformed \stackoverflow{} from a mere Q\&A site into a community project that
gives users a sense of belonging which not only
generates high quality knowledge but also trust in the content that is accumulated.
To emphasize these traits of the content on \stackoverflow{}
that goes beyond questions,
we refer to the content on \stackoverflow{} not as questions but as \textit{posts}.}{3.1}

Before taking part in the community by asking a question for the first time,
users can take a virtual tour that explicates the goals of \stackoverflow{}.
It is explained here, that \stackoverflow{} is \textit{``all about getting answers.
It's not a discussion forum. There is no chit-chat''}. Furthermore,
it asks users to avoid questions that are primarily opinion-based,
or that are likely to generate discussion\footnote{\href{https://stackoverflow.com/tour}{\faIcon{stack-overflow} stackoverflow.com/tour}}.
The platform's focus to avoid \textit{chit-chat}
is also reflected in what~\citet{vadlamani_studying_2020},
and~\citet{zagalsky_how_2016} identified as
the primary drivers behind contributions.
Going beyond a meta analysis of the platform,
scholars used \stackoverflow{} to investigate various aspects of software engineering,
including for example the analysis of trends~\cite{barua_what_2014, yang_what_2016} or
developers' interests~\cite{lee_github_2017}.
Similar to our aim, the \stackoverflow{} data set has also been used to
investigate challenges of software developers.
Based on the assumption that questions and answers on \stackoverflow{} cover a wide range of issues,
\citet{alshangiti_why_2019} analyzed questions in a mixed method study
to identified challenges of software engineers when developing machine learning applications.

\section{Method}

To investigate the lived experience of practitioners on
\stackoverflow{} we take a qualitative approach that aligns with Hoda's
Socio-Technical Grounded Theory (STGT)~\cite{hoda_socio-technical_2022}.
Acknowledging its
iterative nature, we focus on what Hoda defined as STGT's \textit{Basic Stage}
for data collection and analysis.
We take our initial sample from the \stackoverflow{} data dump, which we analyze
using initial and focused coding while we write memos to constantly compare documents,
codes and emerging categories.
\revised{\revised{%
We then present preliminary hypotheses
and an interpretive theory that
summarizes our findings.
Presenting our findings we motivate for the next iteration of our STGT study
that leads to the collection
of more data (theoretic sampling)
to test and extend our findings.
As \citet{hoda_socio-technical_2022} suggests,
we publish our initial findings to assess the relevance of our work
and to receive feedback from the research community.
Successive rounds of data collection and analysis in future work can then lead
to the development of more mature theories that are valuable for the field.
}{1.4}}{1.3}

Our stance with regard to our
research questions is that the reality of testing practices and the experience of
practitioners in a complex socio-technical environment is highly individual and not
reflected by a \stackoverflow{} post in its entirety.
Within the framework of Hoda's STGT we adopt a subjective, constructivist epistemology.
Therefore, we follow Charmaz's version of constructivist Grounded Theory~\cite{charmaz_constructing_2014}
to provide our interpretation of these complex matters.
Despite our awareness of the limitations that an analysis of non-reactive documents has
as they can only provide thin descriptions that lack contextual cues~\cite{fielding_virtual_2008},
we hypothesize that observation
and thorough investigation of attitudes and sentiments expressed by practitioners in posts
on \stackoverflow{} can yield valuable insights into practice.
Furthermore, we claim that our analysis contributes to a better
understanding of socio-technical dynamics in the context of software testing.

To analyze the \stackoverflow{} dataset for our specific purpose
of investigating the sentiment
associated with software testing,
we first retrieve \stackoverflow{} posts related to testing.
We then use sentiment analysis tools to identify posts that contain
negative and positive sentimental expressions.
\revised{%
The whole process,
which starts with this filtering process
of the \stackoverflow{} data dump \circled{0}
and ends with the construction of
preliminary hypotheses and an interpretive theory \circled{22},
is visualized in \Cref{fig:filter}.}{1.6}
\revised{Grounded theory studies
    usually undergo a phase of piloting
    and study preparation
    as a means to verify that
    the chosen tools like questionnaires
    or interview questions
    are appropriately configured and comprehensible to
    the studies' subjects.
    As our study is only involving the analysis
    of non-interactive documents such a verification
    process is not applicable.
    Study preparation in our
    case is thus limited to the extraction of a subset of posts
    that we take from the \stackoverflow{} data dump and the
    configuration of the sentiment analysis tools
that we use (\circled{1} to \circled{5}).}{1.5}

\subsection{Filtering by tags}\label{section:filter:by:tags}
\begin{figure}[] 
	\centering
        \includegraphics[width=1\textwidth]{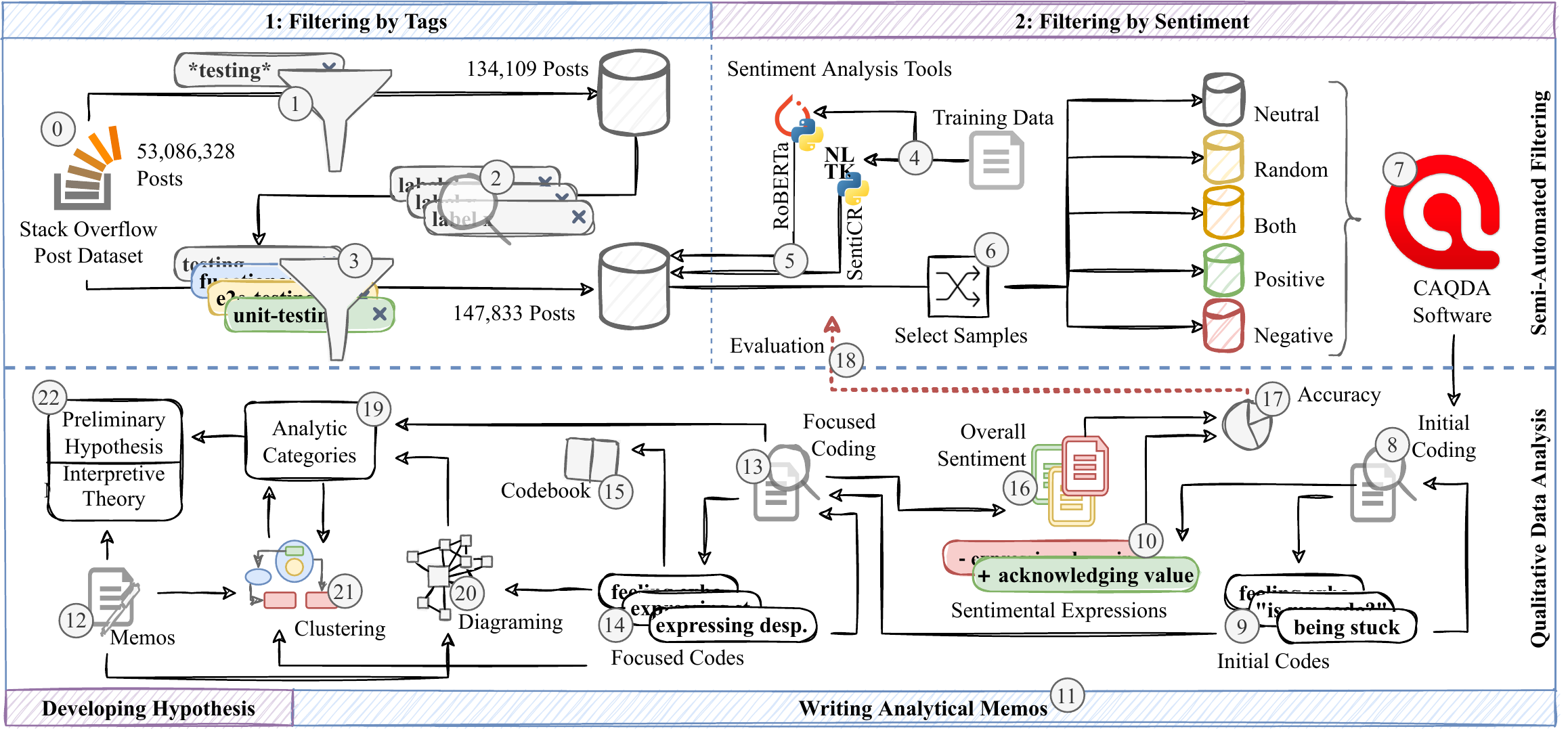}{}
            \caption{\revised{Filtering and annotating \stackoverflow{} posts
                using a semi-automated approach, followed by systematic
        qualitative data analysis process that leads to the construction of preliminary hypotheses and an interpretive theory.}{1.6}}\label{fig:filter}
\end{figure}

The \stackoverflow{} dataset contained 53,086,328 posts concerning all domains of software
development when we obtained it in August 2021
\footnote{%
\href{https://archive.org/details/stackexchange}
{\faIcon{stack-overflow} archive.org/details/stackexchange}}.
To extract a subset with a size that is appropriate for manual analysis,
we filter all posts using a 2-step process
that is outlined in this section. As illustrated in
Figure~\ref{fig:filter}, we begin with the full \stackoverflow{}
\textit{Post}-dataset \circled{0} on the left side and end this process with
importing post-documents into a
CAQDA-software\footnote{CAQDA = Computer-Assisted Qualitative
Data Analysis software; we have mostly used ATLAS.TI,
see: \url{https://atlasti.com}} \circled{7} on the right side.
\revised{%
To extract posts related to automated software testing,
we first filter the dataset using tags.
One or more tags are assigned to every post by their authors.
The list of tags is then often edited by moderators
to facilitate categorization.
Tags represent categories that among others include general concepts or methods
(e.g., \textsc{testing}, \textsc{tdd}),
technologies like programming languages (e.g., \textsc{java}, \textsc{python}),
or specific frameworks and tools (e.g., \textsc{codecov}, \textsc{mockito}, \textsc{reactjs}).
Posts are usually tagged with multiple,
complementary tags
(e.g., post \citedata[]{878848}
is tagged with 5 tags:
\textsc{java},
\textsc{unit-testing},
\textsc{ibdc},
\textsc{mocking} and \textsc{resultset}).
Similar to \citet{yang_what_2016},
we utilized a two-step process
to extract posts by searching for tags which
represent general concepts and methods related to software testing.
We first select all posts from the dataset
that are assigned a tag that contains the word \textsc{testing},
which produces a set of 134,109 posts \circled{1}.
We choose the term \textit{testing}
as it is used as a suggestion on the \stackoverflow{} platform
whenever the tag \textsc{test} is used
and because the tags \textsc{testing} and \textsc{unit-testing}
are the two most prominent tags
when searching for \textit{test} using the tag-search.\footnote{%
\href{https://stackoverflow.com/tags}
{\faIcon{stack-overflow} stackoverflow.com/tags}}
We then manually analyze the list of 13,006 tags
that were assigned to those posts and
remove tags that were used less than 6 times,
or were not directly referring to general concepts of
automated software testing \circled{2}.
The tag \textsc{codecov} for example was
removed from the list because it only occurred 5 times,
and \textsc{reactjs} was removed as it relates
to a programming framework that is not directly related to automated software testing.
We also exclude tags that are related to testing but
focus on a particular technology or tool (e.g., \textsc{mockito}), as we try to
remain testing tool- and development language agnostic.
Following this procedure, we have produced a list of 30 tags that all refer to
conceptual aspects of automated software testing,
like \textsc{unit-test}, \textsc{mocking}, or \textsc{tdd}.
Using this list we again extracted posts from the original dataset.}{1.7}
\revised{%
We extract all posts that contain
at least one tag
that is present on the tag list and
obtain a set
of 147.833 posts \circled{3}.
Post \citedata[]{878848}
which is tagged with
\textsc{java},
\textsc{unit-testing},
\textsc{ibdc},
\textsc{mocking} and \textsc{resultset} was for example selected because
the presence of tags \textsc{mocking} and \textsc{unit-testing}.}{3.3}
We provide the source code of the program that we used to filter posts and the filtered dataset
in our replication package~\cite[filter-by-tags.zip]{swillus_replication_2022}.

\subsection{Filtering by sentiment}\label{section:method_sentiment}
We aimed to examine posts deeply instead of quantitatively which limits our investigation
to an analysis of a small subset of the 147,833 posts.
From research done by other authors we know that only a small fraction of content posted on
\stackoverflow{} contains strong opinions and emotional statements as they mostly discuss
how to use a piece of technology~\cite[]{lin_sentiment_2018}.
Sengupta et al. report that only every 10th comment on \stackoverflow{} expresses some
standalone form of emotion~\cite[]{sengupta_learning_2020}.
This motivated us to create an emotionally rich subset by filtering the dataset using a
semi-automated approach that employs sentiment analysis to select posts that
contain sentimental expressions.
Following the advice of Zhang et al.~\cite[]{zhang_sentiment_2020} to not rely on a single
tool we used the transformer model RoBERTa~\cite[]{liu_roberta_2019} in combination
with the SentiCR tool~\cite[]{ahmed_senticr_2017}.
We trained both tools with a labeled dataset of \stackoverflow{}
provided by~\citet[]{lin_sentiment_2018}~\circled{4}\footnote{Replication package from
\citet{lin_sentiment_2018} containing training data:
\href{https://sentiment-se.github.io/replication.zip}{\faIcon{file-archive} https://sentiment-se.github.io/replication.zip}}.
Their dataset contains
1,500 sentences from \stackoverflow{} posts discussing Java libraries which were
manually labeled by the authors with sentiment
polarities \textit{positive},
\textit{negative}
and \textit{neutral}~\cite[]{lin_sentiment_2018}.
We then used the trained tools,
to automatically annotate sentiment polarities to
every paragraph of every post of our tag-filtered dataset~\circled{5}.
From this annotated dataset we then randomly
extracted posts from 5 categories,
using a simple condition for each category \circled{6}.
\begin{description}
    \item \textbf{Positive:} both tools classified at least one paragraph as positive
        and none as negative
    \item \textbf{Negative:} both tools classified at least one paragraph as negative
        and none as positive
    \item \textbf{Both:} both tools classified at least one paragraph as positive
        and at least one as negative
    \item \textbf{Neutral:} both tools classified all paragraphs as neutral
    \item \textbf{Random:} randomly selected independent of classification
\end{description}

Especially because of concerns raised by Lin et al.~\cite[]{lin_sentiment_2018} and
Jongeling et al.~\cite{jongeling_negative_2017} who state that sentiment analysis
tools often do not provide good results for software engineering texts, we
used the last two categories \textit{Neutral} and \textit{Random} in a later stage of our
analysis to validate our semi-automated filtering approach.
\revised{%
We evaluate whether filtering posts with the tools RoBERTa and SentiCR
provides a dataset
with more sentimental posts
than a random selection.}{1.10}
\revised{%
We choose paragraphs instead of finer grained sentence-level separation
because we hypothesise
that a paragraph is more likely to hold a comprehensive and conclusive thought
as compared to short sentences that are taken out of context.
We argue that sentiment classification done on that level
better supports our goal to group posts into categories of positive and negative posts.
Contrarily to what we want to achieve,
one short and slightly negative remark in a post
of an otherwise very positive paragraph,
is much more likely to determine a wrong result
in a finer grained sentence-level classification.
The sentiment analysis tools we used in this study
both support the approach of classifying text with multiple sentences.}{1.8}
The posts obtained by our semi-automated filtering approach were
imported into a CAQDA software \circled{7} that was used to aid all
further steps of the data analysis.
\revisedtwo{%
To avoid bias during our manual assessment of a post's sentiment,
we did not include the tool's classification result
in those imported posts.
Automatically assigned sentiment was not visible to the authors during manual analysis.}{3.1}
\revised{%
Initially we analyzed 25 posts from each category
(\textit{Random}, \textit{Neutral}, \textit{Positive}, \textit{Negative} and \textit{Both}).
We then added another 25 posts
from each sentimental category
(\textit{Positive}, \textit{Negative} and \textit{Both}),
to reach a point at which the analysis of additional posts did not
provide new insights or perspectives in the form of new codes.
After adding the second batch of 75 posts,
and before reaching the 200th post
we reached saturation.
Posts did not provide new content that did not fit into
the categories which had emerged already at this point.
We therefore analyzed a total amount of 200 posts.}{1.9}
\revised{%
\Cref{fig:top_10_labels} shows
the 20-most occurring tags
that were assigned by authors and moderators to those 200 posts.
When creating our dataset and selecting the posts,
we looked for sentimental discussions about testing
without selecting or excluding specific technologies.
We do not focus on how practitioners sentimentally evaluate
specific tools, e.g., the Java unit testing library \textit{junit}.
We instead take a broader,
tool agnostic perspective.
Nevertheless, to provide context to our dataset,
it is interesting to observe
which tags (both tool agnostic and tool specific)
are assigned to the questions
that are included in our dataset.
In particular, these tags indicate that our dataset
transcends a particular programming language or technology stack.
The replication package we provide contains the source code of our implementation
of the sentiment analysis pipeline~\cite[filter-by-sentiment.zip]{swillus_replication_2022}}{3.3}.

\begin{figure} 
	\centering
            \includegraphics[width=\textwidth]{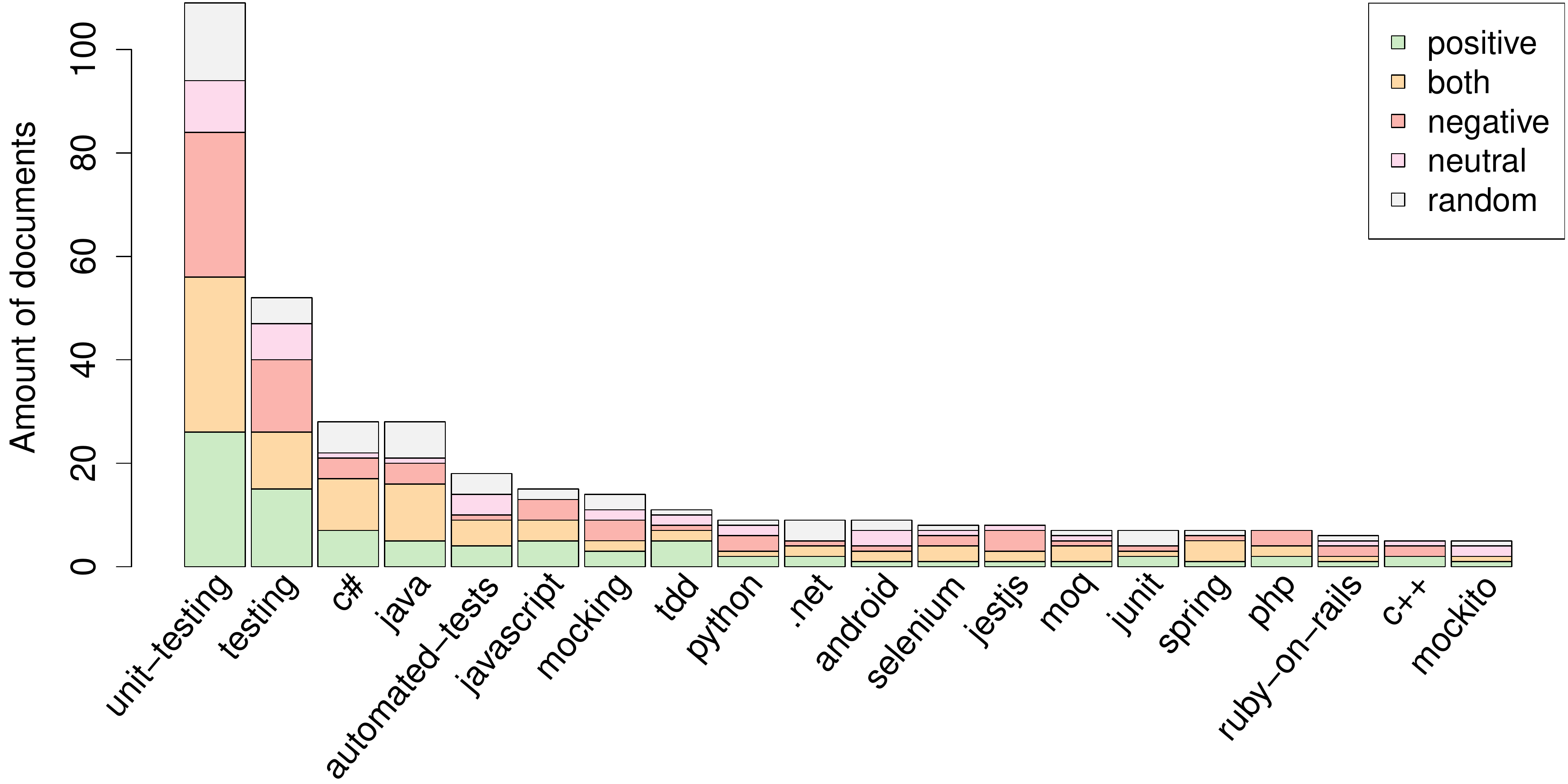}
            \caption{
                \revised{20 most occurring tags of the 200 posts we analyzed.
                The most occurring tags include technology-agnostic tags
                like \textsc{testing}
                and technology specific tags like \textsc{junit}.
                Technology specific tags
                are assigned to posts complementary
                to the 30 more general tool agnostic tags
                that we selected during the filtering process.}{3.3}}%
	\label{fig:top_10_labels}
\end{figure}

\subsection{Data Analysis}\label{section:data:analysis}
\revised{%
We employed strategies from grounded theory as recommended by Hoda~\cite{hoda_socio-technical_2022}
and Charmaz~\cite{charmaz_constructing_2014} to analyze the
filtered \stackoverflow{} dataset.
To begin the iterative process of
constructing abstract analytic categories
out of which we formulated preliminary hypotheses
as illustrated in Figure~\ref{fig:filter},
we use initial coding~\circled{8},
applying codes to the dataset line by line in three rounds.
We started without any preliminary codes, remaining
open to all possible
theoretical directions
especially during the first coding cycle.
In addition to coding posts with gerunds
(e.g., \textit{describing} instead of \textit{description}),
we use \textit{In-Vivo} codes,
which are quotations of what the author of a post wrote in their
own language.
In-Vivo codes are put in between quotation marks and used whenever
authors express themselves in a strong and emotionally rich way
(e.g., \textsc{``is my code just bad?"})~\circled{9}.
\revisedtwo{%
In order to provide basic statistical information about the occurrences of negative and positive sentiment in the dataset, we use
\textit{magnitude coding} as suggested by Saldaña~\cite{saldana_coding_2013}, adding
the symbols \textsc{+} and \textsc{\--} to codes where applicable~\circled{10}.
Negative expressions are coded with a minus
(e.g., \textsc{\==Reflecting unclean approach}), and positive sentiments with a plus
(e.g., \textsc{+Embracing change}) respectively.}{3.1}
We write memos
during all stages
of our data analysis~\circled{11}
which we use at a later stage
to develop preliminary hypotheses~\circled{12}.}{1.11}

After three rounds of initial coding, we reassess the significance of all codes
to decide which ones contribute most to an incise and complete categorization.
As Charmaz suggests, we use this technique to condense the work of the initial
coding phase to advance the theoretical direction of the work
and to begin with a second cycle of focused coding~\circled{13}~\cite{charmaz_constructing_2014}.
During focused coding cycles
we develop focus codes \circled{14} and categorize documents
while we construct and continuously refine a codebook~\circled{15}.
In our codebook we
spell out details like
inclusion- and exclusion criteria,
descriptions, and examples for each focused code.
Because of suggestions made by~\citet{lopez_anatomy_2019},
who have shown that comments on \stackoverflow{}
\revised{can reveal expressions of pride and emotional involvement,
we also incorporate comments made on \stackoverflow{} into our analysis.
Other additional information obtainable via the \stackoverflow{} website,
like the history of changes made by the original author or a moderator
are also considered during focused coding.
We understand a post as a potential entryway
into a deeper and richer context of an author's question.
Details including the sentimental activity in comments,
the editing history of a post both by the author and moderators,
the reasons for a moderator to close a post,
the time it took the community to answer the question
or the fact that it was never answered.
Where a post offers these details (not all of them do),
we capture the information by writing analytical memos.
One memo about post \citedata[]{55357595} with the title \textit{Fruitless pursuit}
written by one of the authors for example reads:}{1.11}
\begin{textbox}[\faIcon{file-signature} Memo: \textit{Fruitless Pursuit}]{}
\revised{%
    \textit{The author of this post did not receive
    any feedback from the community.
    But almost a month after posting
    this question, the author just comments:
    ``Ended up setting up a webpack from the ground up''
    Which I think indicates that this person has
    gone through quite some torment.
    However, they do not express this explicitly.}}{3.1}
\end{textbox}
\revised{%
\revisedtwo{%
During the process of focused coding,
we also assign a sentiment
of positive, negative, both, or neutral
to each post.
Here the assigned sentiment represents
the overall attitude of the author towards testing practices~\circled{16}.}{3.1}
We use both the coding of sentiment~\circled{10}
and assignment of the overall sentiment~\circled{16},
to determine the accuracy of the sentiment analysis pipeline~\circled{17}
and to evaluate its use in our filtering process~\circled{18}.
During the focused coding cycles,
preliminary analytic categories became visible to us~\circled{19}.
A large amount of negative posts containing expressions of desperation for example,
developed into the category \textit{Discouragement} early on.
We refine categories that become visible through the process of coding,
using a diagramming technique described by~\citet{saldana_coding_2013}~\circled{20}.
Starting with a code like \textit{Expressing desperation} or a
post that creates ambiguity when assigned with a category,
we sketch a network of connections to other posts,
categories or codes on paper to explore detailed features of
the coded dataset from different angles.
We then use the clustering strategy as described by~\citet{charmaz_constructing_2014},
grouping posts together and writing memos, concentrating on
commonalities and differences among those groups of posts~\circled{21}.
Taking a different perspective each time,
we find different explanations
for the meaning and context of sentiment
expressed by practitioners in posts.
We continue the process of analyzing the dataset using these strategies,
until they no longer yield new perspectives
and we were able to formulate preliminary hypothesis and an interpretive theory
that emerged from the process~\circled{22}.}{1.11}

\subsection{Constructing Interpretive Theory}\label{section:data:interpretivetheory}
\revised{%
Synthesizing the insights and hypothesis we obtained by engaging
with the data through the whole data analysis process described above,
we formulate an interpretive theory.
Interpretive theory aims to offer accounts
for what is happening,
how it arises
and explains why it happens~\cite[p. 230]{charmaz_constructing_2014}.
In this work we approach interpretive theory
and its construction from a pragmatist viewpoint.
We recognize that our statements can only correlate
our interpretation of the experience of individuals
with our own experience,
and the body of knowledge from the field that is
available and known to us~\cite{mead_mind_2015}.
Taking this viewpoint we emphasize practice and action
rather than trying to explain the empirical phenomena
described in the analyzed data by providing laws that are testable
by empirical objective observation.
Concretely, interpretive theory in this paper
concerns what authors of posts assume about what they describe,
how these assumptions or views might have been constructed, and
how the authors seem to act on their views.
By taking this approach of theory construction,
we want to make phenomena and relationships between them visible
in order to open up new vantage points
for our own and the future work of others.
We understand theorizing as an ongoing activity that can be continued
through this future work~\cite{charmaz_constructing_2014}.}{1.2}

\section{Results}
\revised{%
In this section we describe our findings and offer an interpretation
of the data we analyze to answer the research questions.}{1.2}
We first discuss the result of applying sentiment analysis tools
to create a dataset that is rich in sentimental expression.
We then present the results
of our qualitative data analysis
of this dataset
to first show
how software engineers express sentiment about testing
and which underlying factors contribute to their sentiment.
\revised{%
We then present a preliminary interpretive theory
that synthesizes our findings.}{1.2}
The data in which this preliminary theory is grounded,
and all artefacts that are discussed in this section are
contained in our replication package~\cite[coded-dataset.qdpx]{swillus_replication_2022}.

We invite the reader to import the dataset contained in the replication package in
the CAQDA-Software of their choice,
and we also want to invite the reader to
follow our analysis by using the online content on \stackoverflow{}.
We enable this
by providing a link
to the original post
on the \stackoverflow{} website
that can be followed by clicking on the ID
next to the quotation of a post.
Example quotation: \citedata[This is all working as I would expect]{3340677}.

\subsection{Sentiment analysis for qualitative research}\label{section:evaluation_of_method}

\begin{figure}
	\centering
	    \includegraphics[width=\textwidth]{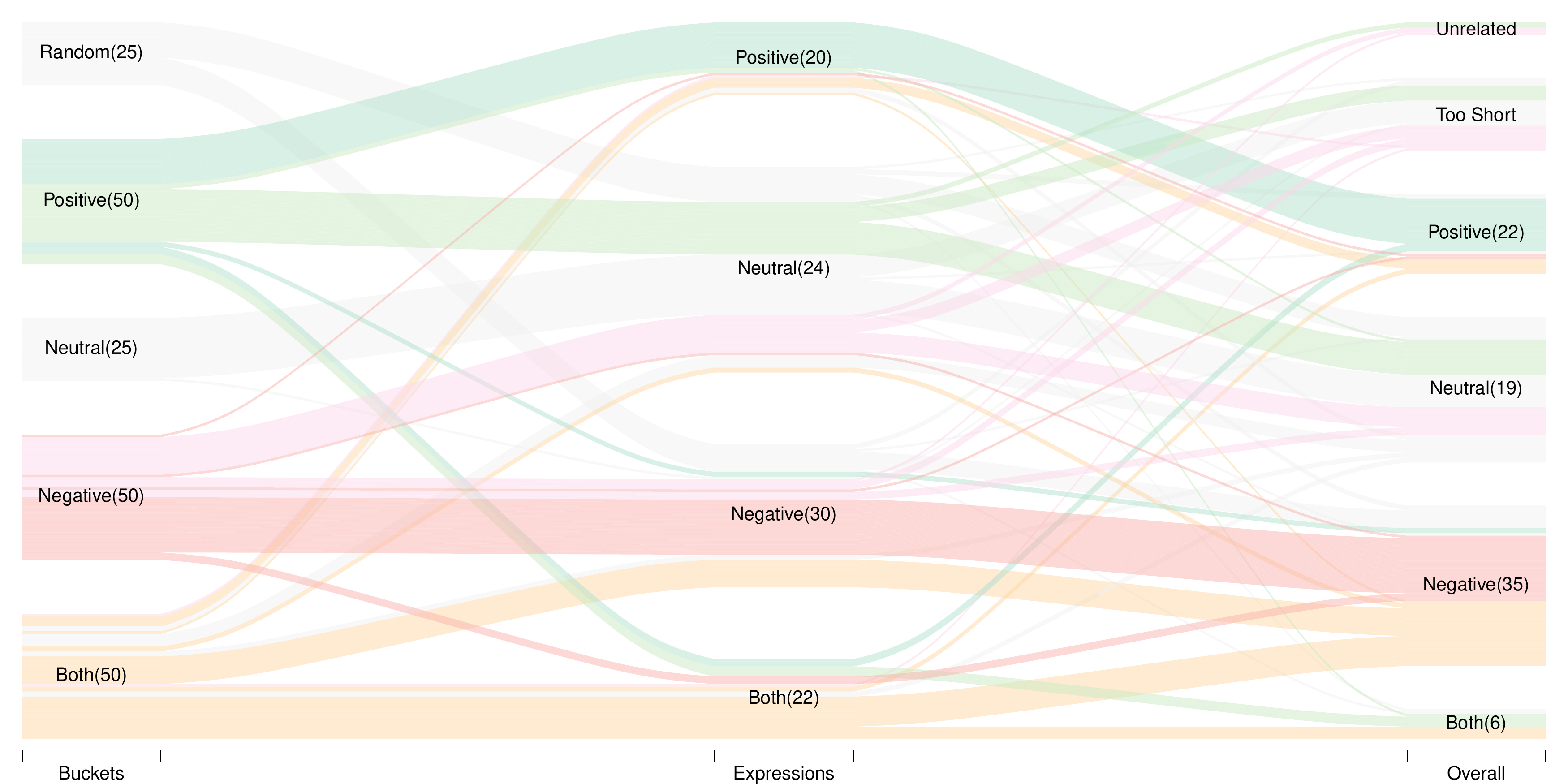}
            \caption{\revised{Visualisation of flows that display how our manual classification of
            posts (column two and three) matches with the automatic classification of our sentiment
            analysis pipeline (column one). The numbers in brackets in column two and three
            indicate the agreement of our manual classification with the automatic classification (first column)}{3.8}}\label{fig:alluvial}
\end{figure}

Our sentiment analysis pipeline takes a \stackoverflow{} post as its
input, classifies each paragraph of the post independently using two
different sentiment analysis tools and takes the result of both tools into account to
indicate if a post is likely to be positive, negative, neutral, or mixed in sentiment.
Using this pipeline we created buckets of positive, negative and mixed sentiment posts,
containing 50 documents each and added 25 neutral and 25 randomly selected posts
to our analysis in order to validate our method.
Our motivation to filter the dataset using sentiment analysis tools stems
from research by \citet{sengupta_learning_2020}, which indicates that randomly selecting
posts from the \stackoverflow{} dataset will only provide few sentimental posts,
as the majority of posts is objective or focused on technical issues.
Our approach relies on multiple
sentiment analysis tools to address a problem that was identified by~\citet{lin_sentiment_2018}
demonstrating that sentiment analysis
can introduce a strong bias when relying on a single tool.
In \Cref{fig:alluvial} we compare
the classification of our sentiment analysis pipeline (left column)
with the sentiment that
we \textit{actually} identified in posts
during initial coding (center and right column).
We differentiate between
occurrences of sentimental expressions in documents (center column)
and the overall sentiment of a document (right column).
Using the metrics
which are visualized in \Cref{fig:alluvial}
we evaluate,
how suitable our method is
to create a dataset
that can be used to
find answers
for our research questions
and if it is applicable for other qualitative studies on \stackoverflow{}.

\subsubsection{Occurrences of sentimental expressions in posts}

Occurrences of sentimental \textit{expressions} in posts were identified and annotated during
the first coding cycle when posts were coded line by line.
\revisedtwo{%
The line \citedata[I understand that using aunit can be a time-saver]{3412892}
was classified as positive for example, but the same post also contains the expression
\citedata[I looked at the aunit manual and I didn't find easy examples to start with]{},
which was classified as negative. Post~\citedata[]{3412892}, which we took from the \textit{positive} bucket,
was therefore assigned the category of \textit{both} sentiments at the level of expressions.}{3.1}
The flow from the first to the second column in Figure~\ref{fig:alluvial}
shows this relation, presenting which posts from each of the sample buckets
contained expressions of the respective sentiment.
20 posts from the bucket of positive posts for example indeed contained one or more
positive sentimental expressions and no negative ones.
In Figure~\ref{fig:alluvial} this relation is represented by the flow from \textit{positive}
in column one to \textit{positive} in column two, highlighted in green.
However, 2 of the 50 posts from the same bucket did not contain a positive expression
but at least one negative expression (flow from positive to negative), 7 posts
contained at least one expression of each sentiment (flow from positive to both)
and 21 posts from the positive bucket did not contain any
sentimental expressions (flow from positive to neutral).
Flows from the \textit{negative} and \textit{positive} buckets to the \textit{neutral}
category in column two indicate that a lot of posts identified as positive or negative
by our pipeline in fact did not contain any sentimental expressions.
Comparing this lack of accuracy with the results for documents that we obtained from the
\textit{random} bucket suggests however that our sentiment analysis pipeline indeed
managed to select more sentimental posts than a random selection would have.
Crucially, we did not find a single positive expression in the set of 25 randomly selected
posts. Additionally, comparing the remaining flows between column one and two in
Figure~\ref{fig:alluvial}, we see that the majority of posts that turned out to contain
sentimental expressions were indeed extracted from the respective bucket.
The findings of this first analysis of the accuracy of the sentiment analysis pipeline
therefore supports our hypothesis that a semi-automated approach proves beneficial
when used to create and analyze a subset of \stackoverflow{} posts with both
negative and positive sentiment.

\subsubsection{Overall sentiment of posts}

In Figure~\ref{fig:alluvial} the last column shows the conformity or difference of the
overall sentiment of posts determined manually by us in comparison with our tool pipeline.
\revisedtwo{%
We determined the overall sentiment of a user towards software testing during the second,
focused coding cycle
and assigned a polarity of \textit{neutral, positive, negative}, or \textit{both}
to each post.}{3.1}
During this analysis, we realized that 39 posts were not usable for
further inquiry.
The majority of those posts were too short (34); one author simply asks
\citedata[Which is the best framework for automatic testing in octave? Why?]{2073244}.
The other five of those unusable posts were identified as unrelated to our work, like
a post in which a practitioner asks \citedata[How to use Jquery Ajax Cache]{2398092},
mentioning testing but referring to something that is unrelated to automated testing.
The dark green and dark red flows in Figure~\ref{fig:alluvial}, from column one via
column two to column three show that
posts from the positive and negative buckets that contain expressions
with that sentiment were mostly leaning into that direction overall as well.
There are only a few outliers of posts
that were for example classified as negative by our pipeline and indeed only contained
negative expressions but were found to express an overall positive sentiment.
One such post contains the \textit{negative} expression that
\citedata[{[it]} is copy-paste code, which I thought was generally not recommended]{9271925},
not mentioning anything positive or negative apart from that. However, the overall sentiment
of the post was interpreted as positive as the author shows a constructive willingness
to improve while being open and concious of their own mistakes.
In total, there were only 12 such cases where the sentiment classification of the
pipeline completely diverged from our classification.
\revisedtwo{%
Documents from the \textit{both} bucket of our dataset, even when they indeed contained
expressions of both sentiments were in most cases negative overall. The analysis also shows
that the \textit{both} bucket contributed the most sentimental posts to our dataset.
Our analysis of the overall sentiment of posts indicates that
subtle remarks and the context of a sentimental expression
makes the overall classification of posts difficult.}{3.1}
Subtracting unrelated (5) posts,
randomly selected posts~(25)
and those that were too short for analysis (34),
we can report
that the sentiment prediction
was correct for 46\% of all documents (65 of 141).
Overall our approach yielded a dataset
in which approximately half of all documents
were sentimental (108 of 200).
We provide an annotation file with our replication package that contains sentiment
annotations for each post that we analyzed on both the level of expression and overall,
including the source code to generate graphs and statistics from that annotation file~\cite[data/annotations.json]{swillus_replication_2022}.

\subsection{Sentiments that affect attitudes}\label{section:coding_results}

\revised{Before describing and comparing occurrences of sentimental expressions
which we identify in the dataset
by presenting focused codes
and analytical categories,
we provide examples which demonstrate
how we moved from the data,
through codes,
towards a more abstract interpretive theory.
Document \citedata[]{878848} was first coded line by line
and was assigned,
among others,
the initial code \textsc{\==Expecting a lot of Work From Mocking}.
The code with the prefix ``\==", which indicates that the expression
reflects negative sentiment,
was assigned to the following line:
\textit{``Use EasyMock, write looooong mocking sequence.
VERY BAD solution: hard to add initial data,
hard to change data,
big test debugging promices."}.
During the second and third initial coding cycle the code
was then changed to \textsc{\==Expecting Mocking to be Bad Solution}.
Other posts hold similar notions
and were coded with the same code
(e.g., \citedata[There is no point
in mocking out a whole ngrx entity store,
so I would just like the selector
to return exactly that object and be done with it.]{58840818}).
During focused coding, the code changed
once again and became more abstract and analytical:
\textsc{``Judging subjectively"}.
The comparison of posts with similar codes revealed
that expectations which are expressed sentimentally,
like the examples above,
are not based on objective observations but
on subjective perceptions
often connected to personal experience.
The intention (or action) of the author here does not seem to be
the objective revelation of their expectations,
but the subjective judgement in order to position themselves.
In one memo titled \textit{Experienced ambiguity}
this notion of subjective judgement and ambiguity
was noted by one of the authors during focused coding.}{1.6}
\newpage
\begin{textbox}[\faIcon{file-signature} Memo: \textit{Experienced ambiguity}]{}
\revised{\textit{The practitioner is struggling
with adopting a new framework.
Some things are easy and some are challenging.
The practitioner is faced with a situation
in which there is no easy or obvious way forward.
They are stuck
and forced to make an uncomfortable decision.
However the willingness to resolve the ambiguity here
still reflects a very positive attitude.
The practitioner already has some clues and they are reasoning from experience.
Looking at the comments, I realized that the post was closed quite quickly.
It only took about 10 hours and the issue was solved by a maintainer
of the framework project which is mentioned in the post.
The fact that the author of the post reacts
very enthusiastically supports my hunch
that their attitude was actually quite positive all along.}}{1.6}
\end{textbox}

\revised{The memo was originally created when analyzing another post (\citedata[]{823276}),
but was then connected to post \citedata[]{878848} as well.
Later, during a diagramming session, the aforementioned memo,
some related focused codes and both posts
(\citedata[]{823276} and \citedata[]{878848})
were assigned to a collection labeled \textit{Confidence}
which generated new memos and more
abstract perspectives.
Both this collection and the memo mentioned above
also contributed to the forming
of the categories \textit{Aspiration} and \textit{Exploration}.
Post \citedata[]{878848},
which ultimately ended up in the category \textit{Aspiration}
and was categorized to reflect both positive and negative sentiment,
further revealed what might be the conditions
for aspiration to arise
in the context of software testing.
We compared the post
with others of the same category
and identified that knowledge and experience seems to
enable practitioners to stay positive despite being stuck
in situations where there is no obvious way forward.
Concretely, we hypothesize that the notion of explicitly
comparing capabilities of approaches,
not only in terms of features,
but also in terms of maintainability,
indicates confidence and experience of the author on \stackoverflow{}.
Ultimately, memos written about those considerations and others
enabled us to construct the preliminary interpretive theory
which we present at the end of this section.
Specifically, the aforementioned post \citedata[]{878848}
supports the hypothesis that experience
and knowledge can give practitioners
an extra degree of trust and confidence,
from which an aspirational attitude
towards testing
seems to emerge.}{1.6}

\subsubsection{Focused codes}
Using focused coding techniques as recommended by \citet{charmaz_constructing_2014},
we identified 22 codes that were assigned to a total of almost 700
different text sections of the 200 posts that we analyzed.
Table~\ref{tbl1} lists all codes, a description for each, and a diagram that
indicates how many posts that contained the code were identified to be either \textit{positive,
negative, neutral} or of \textit{both} sentiments.
The full codebook that we provide as part of the replication package of this paper contains
inclusion and exclusion criteria, and examples for each code~\cite[codebook.ods]{swillus_replication_2022}.

\begin{center}
    \setlength\tabcolsep{0pt} 
    \begin{longtable}{lc}
    \caption{
        Focused codes, their description and the co-occurrences with respective sentiments.
        The co-occurrence bar-chart indicates in how many documents of the overall sentiments
        positive, negative, neutral or both
        the respective code was identified.
        Codes are ordered by the amount
        of documents in which they were found.}
    \label{tbl1} \\
    \toprule
    \textbf{Focused Code (F.X)}  &Sentiment occurrence \\ 
    Description &\fourstackedbarslegend[3.5] \\


    \midrule
    \rowcolor{tablebg}
    \textbf{\fcode{Observing Unexpected Behaviour}{1}} &
    \fourstackedbars[3.35]{1}{0}{41}{20}\\
    \rowcolor{tablebg}
    \descriptionrow{An expression that reveals that something does not work as the author expects.
    Like a dump of error logs that seem to be nonsensical to the author.}\\

    \textbf{\fcode{Reassuring the Reader}{2}} &
    \fourstackedbars[3.35]{13}{4}{26}{18} \\
    \descriptionrow{Making a statement to restore confidence.
    Like a claim that a manual has been read, or a tutorial has been followed.}\\

    \rowcolor{tablebg}
    \textbf{\fcode{Pursuing Ambition}{3}} &
    \fourstackedbars[3.35]{13}{4}{15}{15}\\
    \rowcolor{tablebg}
    \descriptionrow{Constructive attitude to achieve a goal. The implementation of something,
    extension of knowledge or something else
    that goes beyond just getting the job done.}\\

    \textbf{\fcode{Willing to Improve}{4}} &
    \fourstackedbars[3.35]{14}{4}{11}{13} \\
    \descriptionrow{
        Author indicates that they have an ambition to change and improve something.
    }\\

    \rowcolor{tablebg}
    \textbf{\fcode{Facing Uncertainties}{5}} &
    \fourstackedbars[3.35]{10}{9}{16}{8}\\
    \rowcolor{tablebg}
    \descriptionrow{Expression of insecurity through description of ambivalence or doubt.}\\

    \textbf{\fcode{Expressing Desperation}{6}} &
    \fourstackedbars[3.35]{0}{0}{31}{7}\\
    \descriptionrow{Author expresses their desperation directly,
    either by asking a question or by indicating that they are clueless.}\\

    \rowcolor{tablebg}
    \textbf{\fcode{Judging Subjectively}{7}} &
    \fourstackedbars[3.35]{14}{6}{12}{2}\\
    \rowcolor{tablebg}
    \descriptionrow{Explicit subjective valuation of the apparent characteristics, behaviour or value of something.}\\

    \textbf{\fcode{Admitting Lack of Knowledge}{8}} &
    \fourstackedbars[3.35]{5}{6}{16}{9} \\
    \multicolumn{2}{p{\textwidth}}{
        Direct or indirect expression of a lack of knowledge.
    } \\

    \rowcolor{tablebg}
    \textbf{\fcode{Searching for a New Path}{9}} &
    \fourstackedbars[3.35]{10}{5}{5}{14}\\
    \rowcolor{tablebg}
    \descriptionrow{The goal or approach has been thought through
    but the author hunches that there is another, better way.}\\

    \textbf{\fcode{Contemplating Complexity}{10}} &
    \fourstackedbars[3.35]{7}{5}{10}{9}\\
    \descriptionrow{%
     Author is describing something that has to do with the complexity of a setup or use-case.
     Complexity is either highlighted reflected implicitly.
    }\\

    \rowcolor{tablebg}
    \textbf{\fcode{Missing Capability}{11}} &
    \fourstackedbars[3.35]{2}{4}{13}{11}\\
    \rowcolor{tablebg}
    \descriptionrow{
        Description of issues,
        circumstances, hurdles
        or other discomforts
        that stop one
        from reaching a goal.
        Capabilities can be the capabilities of a software,
        its limitations,
        but also the own capabilities to solve an issue.}\\

    \textbf{\fcode{Referring to External Information}{12}} &
    \fourstackedbars[3.35]{8}{3}{10}{8} \\
    \descriptionrow{Reference is made to a resource that is accessible to the author.
        Documentation, blog posts, books etc.}\\

    \rowcolor{tablebg}
    \textbf{\fcode{Contemplating Failure / Difficulties}{13}} &
    \fourstackedbars[3.35]{5}{3}{15}{4}\\
    \rowcolor{tablebg}
    \descriptionrow{%
    Author shares their opinion about what they find difficult or failure they are facing.}\\

    \textbf{\fcode{Looking for Starting Point}{14}} &
    \fourstackedbars[3.35]{7}{1}{11}{8}\\
    \descriptionrow{Request for a starting point to tackle something that is unknown or unclear.}\\

    \rowcolor{tablebg}
    \textbf{\fcode{Facing an Obstacle}{15}} &
    \fourstackedbars[3.35]{1}{1}{14}{3}\\
    \rowcolor{tablebg}
    \descriptionrow{An obstacle makes it impossible to continue with a task.
    The author is stuck because of the obstacle.}\\

    \textbf{\fcode{Reflecting Experience}{16}} &
    \fourstackedbars[3.35]{6}{4}{4}{1} \\
    \descriptionrow{Positive or negative reflection
    which is related to past experience.} \\

    \rowcolor{tablebg}
    \textbf{\fcode{Struggling to Understand}{17}} &
    \fourstackedbars[3.35]{1}{3}{9}{2} \\
    \rowcolor{tablebg}
    \descriptionrow{
    Author is struggling to grasp the meaning of a faced problem
    or a concept they want to learn.
    Like admitting that they are not able to
    comprehend something or that something is hindering them to learn something.} \\

    \textbf{\fcode{Seeing Own Mistakes}{18}} &
    \fourstackedbars[3.35]{4}{1}{3}{7} \\
    \descriptionrow{Realization of an error or a misconception.
    Revelation of having done something
    in the wrong way or in a way that can be improved.} \\

    \rowcolor{tablebg}
    \textbf{\fcode{Comparing Different Approaches}{19}} &
    \fourstackedbars[3.35]{6}{2}{2}{5}\\
    \rowcolor{tablebg}
    \multicolumn{2}{p{\textwidth}}{
        Description of multiple angles to solve an issue or a task.
    }\\

    \textbf{\fcode{Trial and Error}{20}} &
    \fourstackedbars[3.35]{0}{0}{7}{5} \\
    \descriptionrow{Describing different attempts to get to a solution which are all unsuccessful.} \\

    \rowcolor{tablebg}
    \textbf{\fcode{Aiming at a workaround}{21}} &
    \fourstackedbars[3.35]{0}{0}{6}{5}\\
    \rowcolor{tablebg}
    \multicolumn{2}{p{\textwidth}}{
        Practitioner identifies that a situation can be solved by using some workaround
    which is probably not the ideal solution.} \\

    \textbf{\fcode{Excluding Solution}{22}} &
    \fourstackedbars[3.35]{0}{0}{2}{2}\\
    \descriptionrow{There is a solution for a problem but the author does not want or cannot use it.}\\

    \bottomrule
    \end{longtable}
\end{center}

Comparing the codes and corresponding posts
with each other reveals underlying
sentiment of practitioners
that relate to testing practice.
The codes reveal patterns that
affect attitude and testing practices of software engineers and allow
us to propose answers to \textbf{RQ1}.

\begin{textbox}[RQ1]{}
    How do software engineers express sentiment about testing on \stackoverflow{}?
\end{textbox}
\revised{In total, the dataset that we have analyzed
    contains 108 sentimental posts.
    In 32 posts, practitioners expressed positive sentiments,
    63 posts were negative, and 13 contain both sentiments.
    \\
Total amount of sentimental posts: \\
\fourstackedbars[11.5]{32}{13}{63}{0}\\
To highlight some of the patterns
which show how sentiment is expressed,
we elaborate on the eight most occurring codes from \Cref{tbl1}
and explain with examples
what was captured with those codes.}{1.12}

\textbf{Judging Subjectively~\ref{fcode:7}.}
\revised{About one third of sentimental posts (32 of 108)}{1.12}
contained an explicit subjective statement
about apparent characteristics or value.
Subjective expressions like that of one practitioner who
\citedata[fell in love with the crisp syntax {[of a framework]} immediately]{1072952}
underline the attitude of the author.
Negative attitudes connected to judgement
like one practitioner reflecting on a specific practice which
\citedata[seems like a waste of time]{29894788}
\revised{were rarer in the dataset
than positive attitudes.
One practitioner for example reflects positively
\citedata[that {[running tests concurrently]} will force {[them]}
to refactor some code to make it thread-safe,
but {[they]} consider that to be a good thing :-)]{4970907}.
In total, more than one third of all positive posts (14 of 32)
contained a subjective judgement
compared to only every fifth negative post (12 of 63).}{1.12}

\textbf{Lack of Knowledge~\ref{fcode:8}, Facing Uncertainties~\ref{fcode:5} and Reassuring the Reader~\ref{fcode:2}.}
Outlining the limits or lack of their own knowledge and abilities
by stating for example that they are
\citedata[a newbie]{29894788}, or indirectly pointing out that they are
\citedata[stuck trying to {[...]} test an extremely simple project]{62177256}
occurs both in positive and negative posts
\revised{in around a quarter (27 of 108) of all sentimental posts}{1.12}.
In addition to describing their own limits by admitting a lack of knowledge,
we identified descriptions of ambivalence (\citedata[Which is the correct way?]{41262775}),
doubt (\citedata[Has anyone done anything similar before or is this crazy?]{7213917}), or
uncertainty (\citedata[It seems to me that,
I maybe should be creating a Fake MaterialRepository, rather than mocking it?]{23534123})
expressing insecurity
\revised{in around a third (35 of 108) of all sentimental posts.}{1.12}
We also found statements
indicating that the author is trying to maintain or restore
their confidence by reassuring the reader
\revised{in more than a third of sentimental posts (43 of 108)}{1.12}.
One author for example is stuck in a situation where they observe something unexpected
and they \citedata[want to understand why that is like this]{39592949}, wondering if
\citedata[there is a better way]{}, even being afraid that their \citedata[code is just bad]{}
but still holding on to their approach as they reassure the audience that
\citedata[When {[they]} change {[something,]} everything works fine]{}.

\textbf{Pursuing Ambition~\ref{fcode:3} and Willing to Improve~\ref{fcode:4}.}
\emph{Uncertainties} and a \emph{lack of knowledge} were found equally frequent
in negative and positive posts, but
descriptions of constructive attitudes to achieve a goal
that goes beyond just getting the job done
were mostly found in positive posts,
or posts that contain both sentiments.
We identified direct expressions of ambition by practitioners for example
\citedata[to create a support library that could be used by all test projects]{18399610},
or mentioning the context of a challenge that underlines its ambitious nature like
\citedata[writing acceptance tests for a single feature of a large App {[, needing]}
a lot of data for this and {[having]} a lot of scenarios to test]{28129825}.
Those expressions were found
in over a third of positive posts (13 of 32)
but contrary only in around
one fourth of negative posts (15 of 63).
Related, and very similar to these expressions
are verbalized intentions to improve,
for example by wanting to \citedata[structure {[a]} unit test in a better way]{43275116} or
by asking for \citedata[the best practice in {[a particular]} case]{46177956}.
Just like mentions of ambitions,
expressions of a willingness to improve
\revised{%
occurred in more than one third of all positive posts (14 of 32),
and contrarily only in less than a fourth of negative posts (11 of 63).
Together, expressions of ambition and willingness to improve
cover almost three quarters of all positive posts (23 of 32).}{1.12}

\textbf{Expressing Desperation~\ref{fcode:6} and Unexpected Behaviour~\ref{fcode:1}.}
Contrarily to ambitions we also found expressions of despair by practitioners who are
stuck saying that they for example
\citedata[googled wide and far, but did not get any answer]{58840818},
or remain completely helpless, begging for support like one practitioners who asks:
\citedata[Can somebody please, please, please for Pete's sake
{[...]} fix this bug that
thousands are having?]{44762082}.
We did not observe expressions of desperation in positive posts
or posts with both sentiments,
\revised{%
but we did find them in almost half (31 of 63) of negative posts.
Additionally, we identify descriptions of unexpected behavior
in more than half
of negative posts (41 of 63).}{1.12}
Covering a big fraction of the dataset, unexpected behavior
is experienced by practitioners in many different contexts, referring to testing
practices or the development environment (\citedata[When I test it in browser,
everything is OK, because App{\textbackslash{}User} exists, but when I test my plugin,
App{\textbackslash{}User} doesn't exists]{52760148}),
or referring to something that is not directly related to testing but discovered
through it like facing a floating point precision error for the first time, noticing that
\citedata[When I'm running the tests it's broken because 0.1 is not equal to 10{\%}]{63886733}.

\subsubsection{From Codes to Categories}
We use codes to compare posts with each other
in a structured way.
Codes enable us to scrutinize the dataset from different perspectives.
Co-occurrences of codes within posts for example
reveal patterns in the data that can be indicators for categories.
We identified four major factors
that describe the non-technical, situational context
of sentimental posts with which we can categorize the posts.
In this section we present each category and their characteristics,
highlighting key insights that emerged from the data during our analysis
when categories were outlined.
The categories reveal underlying currents that affect
the testing practices of software engineers.
Categories which highlight what influences their attitude and motivation
are the basis of what we propose as answers to \textbf{RQ2}.

\begin{textbox}[RQ2]{key:1}
    Which factors affect sentiment of software engineers towards testing practices?
\end{textbox}
\begin{multicols}{2}
\textbf{\category{Discouragement}{1}} \\
\rightline{\fourstackedbars[5]{0}{0}{42}{10}}
\end{multicols}
\revised{%
We found that attitude in negative sentimental posts is often (42 out of 63)
expressing \textit{discouragement~\ref{cat:1}} from testing.
\textit{Unexpected behavior~\ref{fcode:1}} can bring efforts to a halt
\reflist{6441026}{37439708}{53935108}{44095109}{31052776}{32408965}{55644155}{56577906},
sometimes made explicit in posts
by references to \textit{an obstacle that is faced~\ref{fcode:15}}
\reflist{8338348}{32622060}{20480791}{38932495}{6376925}{6376925}.
\textit{Expressions of despair~\ref{fcode:6}} underline
the weight of these setbacks in those posts
\reflist{3736614}{58840818}{67734277}{53935108}{17068154}{33607092}{44762082}{61782427}.
When authors sentimentally express discouraging setbacks in their testing efforts
by \textit{contemplating difficulties or failure~\ref{fcode:13}}
they are at the same time often \textit{reassuring the reader~\ref{fcode:2}},
implying that the problem cannot be blamed on them
\reflist{63795587}{14942409}{19490583}{18083834}{19799393}{25264248}{26370705}.
Statements that a \textit{tutorial or documentation~\ref{fcode:12}}
was followed and thoroughly read,
or reports of \textit{elaborate debugging~\ref{fcode:20}}
demonstrate the confidence of the author
\reflist{13309278}{32009877}{34889215}{6579379}{14701609}.
A \textit{complex development environment~\ref{fcode:10}},
including company policies or unique infrastructure configuration
is mentioned in the context of such cases
\reflist{6475042}{42211311}{14554366}{18038203}{17068154}{43435227}.
In complex situations,
even a small step,
like writing a unit test,
can cause a lengthy and often fruitless pursuit
\reflist{52760148}{37527179}{55357595}{36608077}{67734277}{14942409}.
When tools, methods, and concepts are \textit{not easily understandable~\ref{fcode:17}},
especially when \textit{documentation is not extensive enough~\ref{fcode:11}},
practitioners are discouraged to hold on to their ambition
\reflist{44010437}{63795587}{61769730}{19799393}{43435227}{7292700}{62177256}.}{1.6}
\medskip
\begin{textbox}[Recapitulation: Emergent Category \textit{Discouragement}]{key:2}
    \begin{minipage}[t]{0.35\textwidth}
      \centering\raisebox{\dimexpr \topskip-\height}{%
      \includegraphics[width=\textwidth]{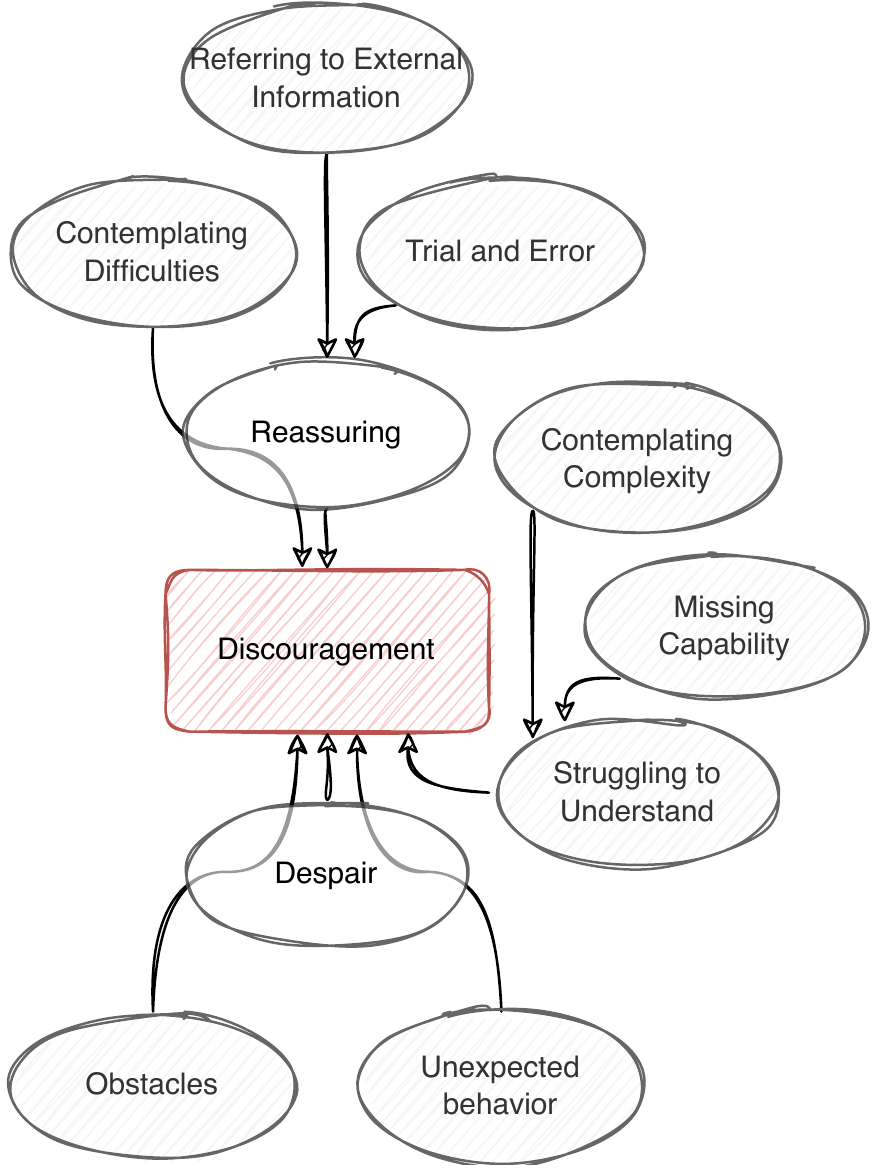}}
      \captionof{figure}{\revised{Focused codes and how they are related to the analytical category \textit{Discouragement}}{1.13} }
      \label{fig1}
\end{minipage}\hfill
\begin{minipage}[t]{0.6\textwidth}
\revised{%
Discouraging sentiment about testing is provoked in
complex development environments.
This includes company policies or unique infrastructure configuration.
When such factors combine with technical issues,
experienced by the practitioner as unexpected behavior,
they create obstacles that discourage practitioners from testing.
A complex environment
makes the usage of a standard testing tool chain
unexpectedly challenging,
especially when practitioners lack experience in testing.
Documentation or other external resources do not help in these cases
and long fruitless pursuits of trial and error are reported.}{1.13}
\end{minipage}
\end{textbox}
\begin{multicols}{2}
\textbf{\category{Exploration}{2}}\\
\rightline{\fourstackedbars{13}{5}{10}{16}}
\end{multicols}

\revised{%
Contrary to posts in which a discouraging sentiment is expressed,
posts of practitioners who approach testing
with an \textit{exploratory~\ref{cat:2}} sentiment,
reflect both positive and negative attitudes.
In the context of exploration,
reaching out to the \stackoverflow{} community
is motivated by an \textit{ambition~\ref{fcode:3}} to overcome
\textit{difficulties or failure~\ref{fcode:13}}.
\reflist{57299238}{29305776}{53376098}{3340677}.
We observe that practitioners have a positive attitude when they
indicate a \textit{willingness to improve\ref{fcode:4}},
\reflist{53657417}{41135403}
especially if they are \textit{searching for a new path~\ref{fcode:9}}
\reflist{14602848}{946069}{6022092}{7213917}
to solve a problem
by asking for available best practices in a particular situation.
\textit{Resources~\ref{fcode:12}} like a blog post or documentation
or other external factors seem to trigger positive ambitions
of practitioners in those cases
\reflist{67709670}{32046670}{4659714}.
When practitioners are struggling with basic concepts however,
for example by \textit{looking for a starting point~\ref{fcode:14}},
they show a negative attitude
\reflist{61342139}{55176792}{3412892}{57609818}.
Exploration with this negative attitude is connected to
acknowledgement of a \textit{lack of knowledge~\ref{fcode:8}}
\reflist{59729159}{46713912},
or \textit{uncertainties~\ref{fcode:5}} about practices.
\reflist{12659810}{7960832}{4288448}{2894608}{823276}.
Crucially, in cases where practitioners that explore testing
report \textit{unexpected behaviour~\ref{fcode:1}},
their attitude is exclusively negative
\reflist{49480999}{37439708}{29305776}.}{1.6}

\medskip

\begin{textbox}[Recapitulation: Emergent Category \textit{Exploration}]{key:2}
\begin{minipage}[t]{0.35\textwidth}
  \centering\raisebox{\dimexpr \topskip-\height}{%
  \includegraphics[width=\textwidth]{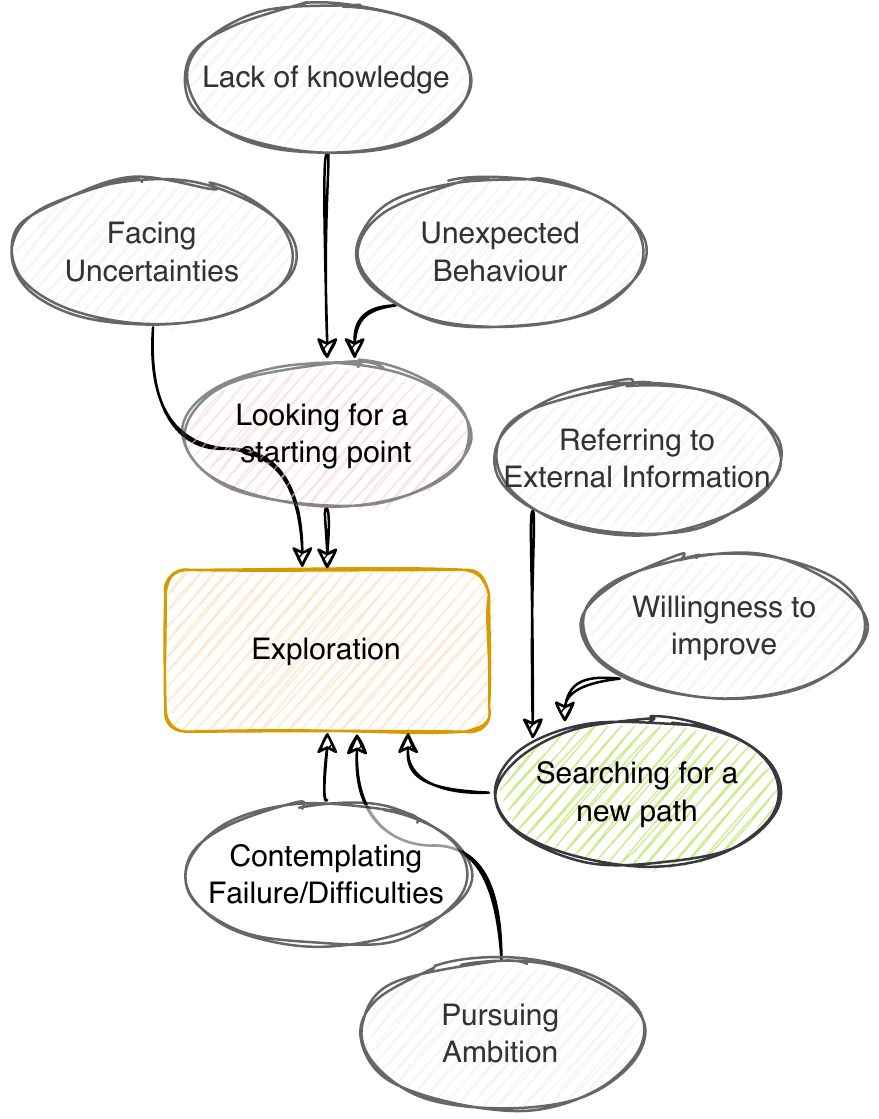}}
  \captionof{figure}{\revised{Focused codes and how they are related to the analytical category \textit{Exploration}}{1.13} }
  \label{fig:exploration}
\end{minipage}\hfill
\begin{minipage}[t]{0.6\textwidth}
  Exploratory sentiment to discover and learn
  is expressed both positively and negatively by practitioners.
  Trust into method or technology based on experience
  or inspiring external impulses
  arouses positive attitudes.
  When exploration serves clarification in situations of uncertainty,
  it is the experience of unexpected behaviour of technology that
  causes negativity especially when practitioners lack experience.
\end{minipage}
\end{textbox}
\newpage
\begin{multicols}{2}
\textbf{\category{Reflection}{3}}\\
\rightline{\fourstackedbars{8}{3}{11}{8}}
\end{multicols}
\revised{%
We identify negative and positive posts in which practitioners
sentimentally and critically \textit{reflect}
on their testing practices or understanding.
\textit{Reflection of experiences~\ref{fcode:16}}
and expressions of an ambition to \textit{improve~\ref{fcode:4}}
when they are \textit{facing uncertainties~\ref{fcode:5}}
form the baseline of this category
\reflist{398004}{59781140}{49713083}.
Similar to the posts we categorized as \textit{exploration},
\textit{uncertainties of practitioners~\ref{fcode:5}} are directly expressed
or indicated through attempts to \textit{reassure the reader~\ref{fcode:2}}
\reflist{41262775}{58684292}{687748}{29894788}.
In this category however, we observe that practitioners are
more \textit{aware of their mistakes~\ref{fcode:18}}
or their \textit{struggle to understand~\ref{fcode:17}} aspects of testing
\reflist{42275344}{4991264}{4970907}{67746901}{39892949}.
Posts that reflect a positive attitude contain analysis and
\textit{comparisons of approaches~\ref{fcode:19}}
\reflist{46177956}{41816229}.
In contrast, when practitioners
\textit{contemplate failure~\ref{fcode:13}} or \textit{complexity~\ref{fcode:10}}
their attitude is almost exclusively negative
\reflist{64464005}{1064403}{18941509}{25325133}{42374231}.}{1.6}
\medskip

\begin{textbox}[Recapitulation: Emergent Category \textit{Reflection}]{key:3}
\begin{minipage}[t]{0.35\textwidth}
  \centering\raisebox{\dimexpr \topskip-\height}{%
  \includegraphics[width=\textwidth]{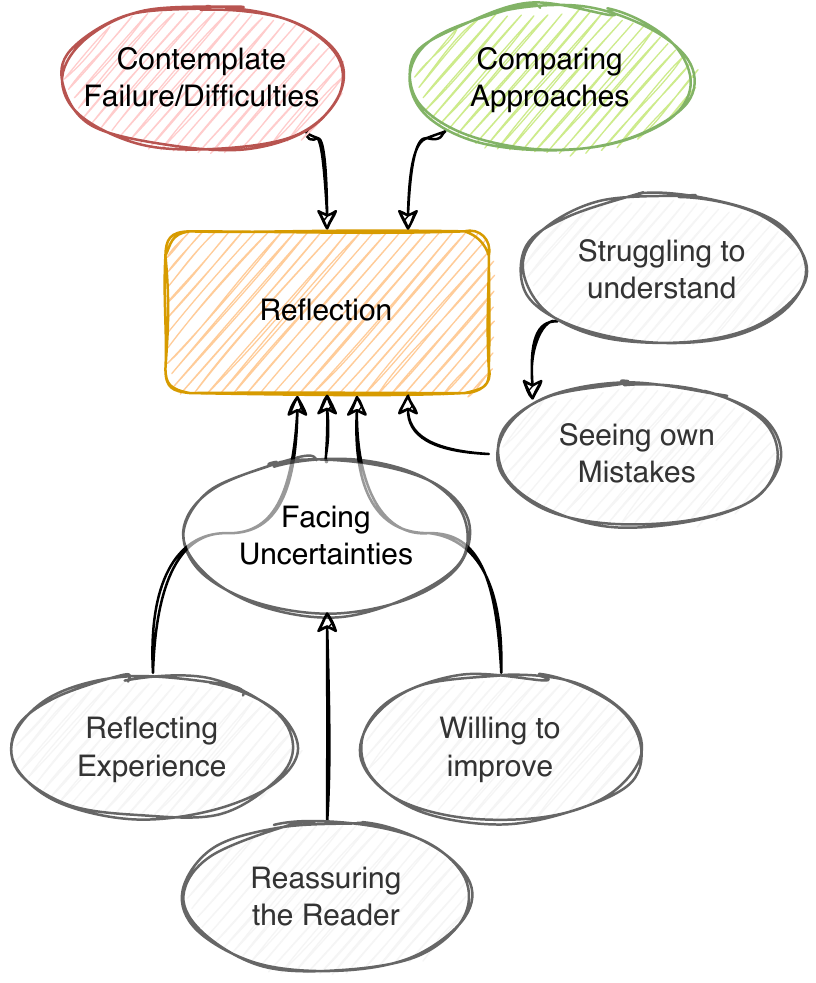}}
  \captionof{figure}{\revised{Focused codes and how they are related to the analytical category \textit{Reflection}}{1.13} }
  \label{fig:exploration}
\end{minipage}\hfill
\begin{minipage}[t]{0.6\textwidth}
    Application of testing practices
    can lead to ambiguity.
    Applying the right method
    in a particular situation for example can be challenging.
    Awareness of blind spots
    and knowledge of the great variety of tools and methods,
    is a factor that allows practitioners
    to keep a positive attitude.
    Variety and ambiguity can than even be appreciated.
    When failure or complications cause ambiguity however,
    sentimental reflection is negative.
\end{minipage}
\end{textbox}
\begin{multicols}{2}
\textbf{\category{Aspiration}{4}} \\
\rightline{\fourstackedbars{11}{3}{0}{5}}
\end{multicols}

\revised{%
Opposite to posts from the category of \textit{discouragement},
we identify \textit{aspiration} in posts
which express almost exclusively positive attitudes towards testing.
Specifically, aspiration reflects a degree of freedom
that allows exploration and discovery in a constructive way.
In particular, the motivation is not to find a workaround or to overcome an obstacle,
nor do authors elaborate on extensive debugging or trial and error.
Instead, authors \textit{pursue ambitions~\ref{fcode:3}}
that go beyond a particular situation
\reflist{34657563}{22246656} and express
\textit{intentions to improve~\ref{fcode:4}} their testing practices
\reflist{16938742}{14961412}{48113464}{6684337}{878848}.
Facing \textit{complex situations \ref{fcode:10}}
is here not a cause for distress,
but rather a motivation to improve testing practices
\reflist{28129825}{1072952}{23062243}.
Motivation is expressed by authors
through explicit positive \textit{judgments of value~\ref{fcode:7}} of testing
\reflist{280645}.
The post on \stackoverflow{} can in those cases be
an attempt to \textit{find a new way~\ref{fcode:9}} to tackle a problem
\reflist{9271925}
or to probe for a \textit{starting point~\ref{fcode:14}}
\reflist{1006189}{52539907}.}{1.6}
\medskip

\begin{textbox}[Recapitulation: Emergent Category \textit{Aspiration}]{key:4}
\begin{minipage}[t]{0.35\textwidth}
  \centering\raisebox{\dimexpr \topskip-\height}{%
  \includegraphics[width=\textwidth]{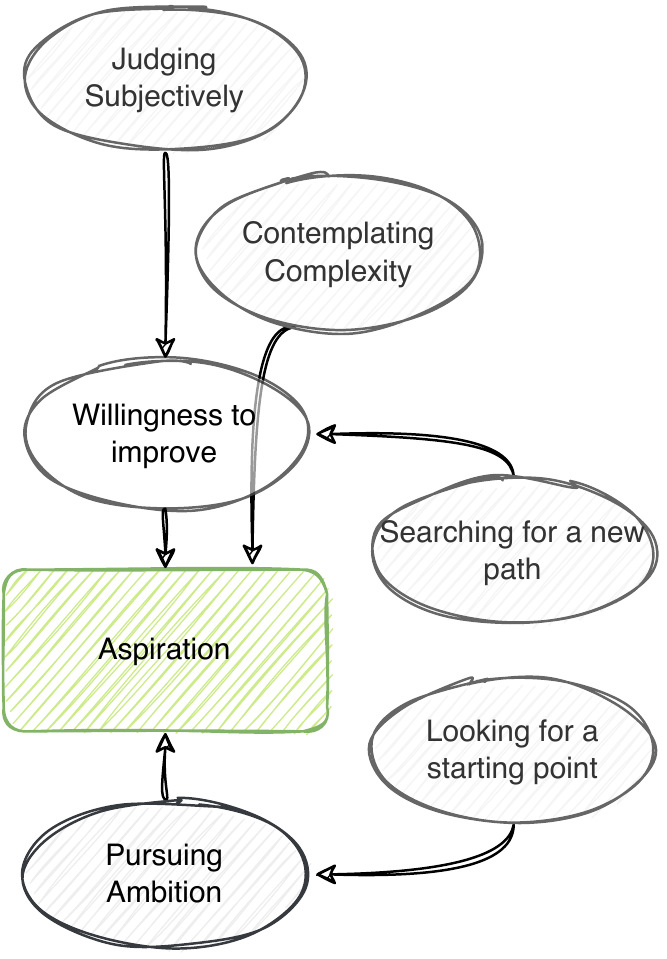}}
  \captionof{figure}{\revised{Focused codes and how they are related to the analytical category \textit{Aspiration}}{1.13} }
  \label{fig:exploration}
\end{minipage}\hfill
\begin{minipage}[t]{0.6\textwidth}
Understanding of long term goals
and the value of testing
arouses aspirational sentiment.
Not being trapped in a problematic or complicated situation
and not having to deal with an immediate obstacle
creates space that is required for this aspirational attitude.
It allows practitioners to build essential knowledge
before their ignorance produces problems.
\end{minipage}
\end{textbox}

\subsubsection{Factors that arouse sentiment}
To answer \textbf{RQ2}
(Which factors affect sentiment of software engineers towards testing practices?),
we summarize key insights we gained
by developing the above categories.
We identify that practitioners on \stackoverflow{}
express sentiments when they are either
discouraged~\ref{cat:1} from pursuing their goal,
aspiring~\ref{cat:4} towards something
that goes beyond their usual practice,
reflect~\ref{cat:3} on their testing experience and knowledge,
or when they are exploring~\ref{cat:2} what is still unfamiliar to them.
Posts which indicate aspiration~\ref{cat:4} are positive in sentiment,
and posts that describe notions of discouragement~\ref{cat:1} from testing
mostly reflect negative sentiment.
Common factors can be identified
even among those two almost inverse categories.
Concretely, we identify that
the experience of unexpected behavior
is an important factor
that leads to negative sentiment
expressed through discouragement.
Even when exploring~\ref{cat:2}
or reflecting on~\ref{cat:3}
testing practices
to learn and gain knowledge
practitioners express negative sentiments
when they face unexpected behavior that causes ambiguity.
Additionally, data suggests
that an absence of those unexpected setbacks
enables conditions for practitioners to aspire.
Through reflection and exploration,
these conditions allow them
to build knowledge and experience.
Experience, which is likely
to prevent those unexpected setbacks
in the future.
Trust in testing practices
that is established through these experiences
contributes to positive sentiments
when new practices are explored.
We find the same to be the case for an awareness of blind spots.
Reflection~\ref{cat:3} on their testing practices
that express an awareness of blind spots
reflects positive sentiment and attitude.
Uncertainty in those cases
inspire practitioners instead of discouraging them.

\subsection{Trust, Complexity and Testing - Preliminary Theory}
\label{section:theory}
We set out to discover
what makes practitioners sentimental about testing
by looking at how they express sentiment on \stackoverflow{}.
We want to know which factors and situations contribute to sentiment.
By analyzing, categorizing, and comparing the dataset,
we got a glimpse of what the
experience of practitioners,
who ask questions on \stackoverflow{} must be like.
Codes and categories described in the previous sections
enabled us to analyze the dataset systematically
using techniques like clustering and diagramming.
In this section we present a preliminary interpretive theory
that describes what became visible
from our perspective,
which is grounded in the analyzed dataset.
To let the data speak for itself,
we provide references to the original posts on \stackoverflow{}
immediately in the text.
With each quotation from posts,
we also provide a reference to
the code that was assigned
to the respective text section
where applicable.
\Cref{fig:theory} illustrates
our preliminary theory as an
interplay of
the most crucial factors
which we identified
to have an influence
on sentiment towards testing on \stackoverflow{}.
We first elaborate on
the right side of the figure,
which shows discouragement~\ref{cat:1} in the context of software testing,
and how the negative sentiment around it is aroused
in situations where
complexity plays a central role.
We then turn to the left side of the figure,
elaborating which role
exploration~\ref{cat:2}, reflection~\ref{cat:3} and aspiration~\ref{cat:4}
play in the context of testing.

\begin{figure}[] 
	\centering
	    \includegraphics[width=1\textwidth]{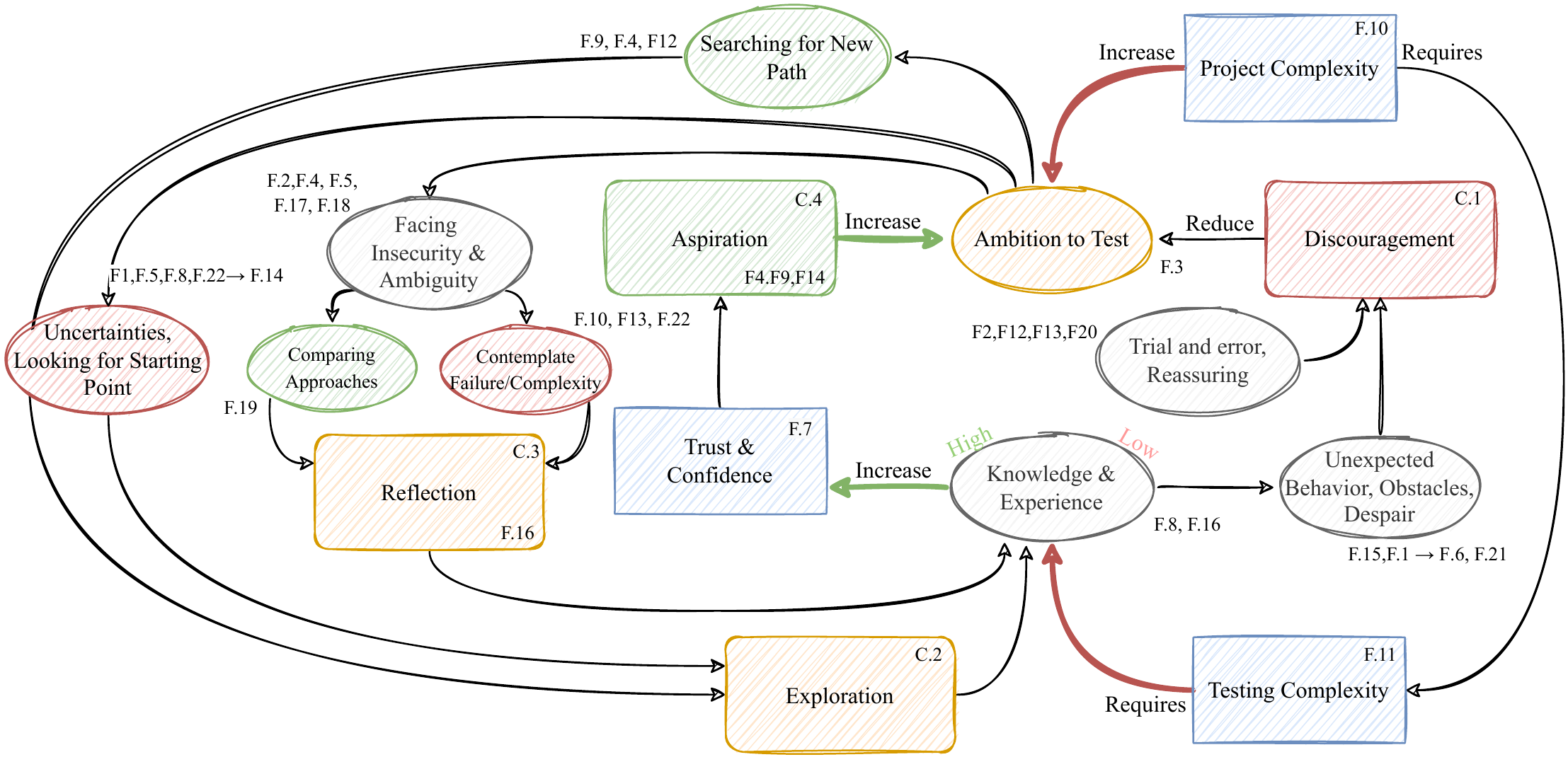}
            \caption{%
                \revised{
Interdependence of factors
which lead to sentiments around testing
and how they are aroused and amplified
in the context of
complexity, trust and confidence through (the lack of) knowledge and experience.}{1.15}}
	\label{fig:theory}
\end{figure}

\textbf{\citedata[I was starting to break as much as I was fixing. So I decided I'll start from scratch, with TDD this time]{29894788}~\ref{fcode:3}.}
Testing practices and approaches are multi faceted.
Even in cases where practitioners are just
\citedata[having a play with testing]{28129825}~\ref{fcode:4} to improve their code base, or just to
\citedata[understand the essence of it]{44202672}~\ref{fcode:3}, they are quickly faced with multiple tools
and have to make difficult choices regarding the technique or tools to adopt for a use-case.
The dataset that we analyzed demonstrates that testing software is not a
single tool or single method practice.
\revised{%
We observe that the big landscape of software testing tools and the resulting
diversity of possibilities to practice testing
amplifies ambivalence when practitioners lack experience
and knowledge \reflist{878848}{1006189}{12950163}{601973}{17320143}.}{1.19}
\revised{The question whether
or not \citedata[I {[am]} missing something in my pursuit of cool and
trendy stuff {[...]} ditching the old proven {[ways]}]{2894608}~\ref{fcode:5}
expresses the lingering
insecurities of practitioners
who are plunging into a world
where many and often unexpected
aspects of software engineering
suddenly come together \reflist{823276}{43435227}{1454949}.}{1.19}
As software projects get more complex, the ambition
\citedata[to fully automate testing {[...]} in the most simple way possible]{16938742}~\ref{fcode:3}
using advanced practices that are able to tackle this increased complexity grows as well.
\revised{Our investigation indicates
that this clash of
lack of experience in testing on the one hand, and complicated challenges on the other hand
drives attitudes around software testing \reflist{4991264}{43435227}.
As shown in \Cref{fig:theory},
as a circular pattern,
we identify that
a growth in complexity of
either the development environment
or the software project itself
makes practitioners ambitious
to learn (more) about software testing \reflist{1072952}{16938742}{1006189}.}{1.19}
But a high level of complexity of production code
(top of \Cref{fig:theory})
also requires complex testing code
which in turn requires more than
basic knowledge of testing (bottom of \Cref{fig:theory}).
The interplay of
growing ambition, a complex environment, and a lack of knowledge
is reflected in a question
about an \textit{easy} way to write a unit test.
The practitioner asks:
\citedata[I'm refactoring one big complicated piece of code {[...]}.
So, I need to write a unit test]{}~\ref{fcode:3}\citedata[{[...]}. After googling I came up with 2 ideas]{}~\ref{fcode:8} \citedata[{[...]}.
Am I missing some silver bullet? Possibly, DBUnit is the tool for this?]{878848}~\ref{fcode:9}.
Unfortunately, practitioners only start
to face their ambiguities and insecurities
around testing
when they are \citedata[starting a new project,
that promises to be much bigger
and more involved
than anything {[they]} have done in the past]{6684337}~\ref{fcode:4}.
\revised{%
In other words:
instead of learning testing practices,
starting with simple comprehensible setups
and then iteratively building knowledge
as the complexity of test suites and source code
under test grow simultaneously,
practitioners throw themselves into cold water
when it is too late
for simple, approachable solutions \reflist{19490583}{6475042}{53657417}{4659714}.
When the \textit{silver bullet} is not found,
they get discouraged to continue with their ambition \reflist{878848}{63795587}{14942409}{7960832}.}{1.19}
Our data analysis suggests
that discouragement~\ref{cat:1} is often connected
to this phenomenon
as expressions of desperation~\ref{fcode:6} indicate
strong negative sentiment when practitioners
are stuck~\ref{fcode:15},
sometimes after they already \citedata[googled wide and far]{58840818}\ref{fcode:6},
\citedata[searching for days to find an answer]{43435227}\ref{fcode:6}.
\revised{%
Unhelpful gathered information~\ref{fcode:12}
which is often referenced in \stackoverflow{} posts
only increase negative sentiment,
and sometimes leads practitioners to identify unexpected
behavior~\ref{fcode:1} of testing tools and libraries as \textit{weird} or
\citedata[strange behavior, because documentation says {[that something should work. But:]}
Well, this is not happening.]{63795587}~\ref{fcode:7}~\reflist{19490583}{26370705}.
An explanation for this could be that
documentation of testing tools and tutorials
for beginners are more likely to focus on
simple and
standard use-cases \reflist{57609818}{6475042}{13309278}{37527179}{34889215}{14701609}.}{1.19}
Based on our anecdotal experience
as software engineers using testing practices,
we hypothesize that
a divergence from best-practices in
both software design and development environment,
requires practitioners to rely on testing experience.
In the context of highly inventive or original approaches,
simple tutorials for testing are not applicable.
It is very likely that
more than one testing library
is required in those complex
non-standard software environments.

\begin{textbox}[Complexity in Testing Practice]{}
Before we set out to investigate
what lies behind sentiment around software testing
on \stackoverflow{},
we assumed that it will
mostly be connected to
tool failure or bugs.
We expected to find sentimental complaints
about specific (missing) features
in a specific version of libraries for example.
Our analysis shows however that
it is more likely to be
a struggle in overcoming overwhelming complexity
with methods or combinations of tools
that practitioners are not experienced enough with
which causes negative sentiment.
\end{textbox}

Testing software can confront practitioners
with misconceptions or flaws
of their software projects.
One practitioners asks:
\citedata[Is this a valid unit test? If not, is it because I have bad design {[...]}?
Because currently, I see absolutely no benefit in writing this test]{44202672}~\ref{fcode:5}.
Even as the majority of sentimental post that we analyzed reveal discouragement and
negativity as described in the preceding paragraphs,
some authors maintain a constructive and
even aspirational attitude~\ref{cat:4},
even when they are facing difficulties~\ref{fcode:15}.
We observe that positive posts
rarely contain descriptions of unexpected behavior or expressions of desperation.
\revised{In contrast,
even in difficult situations,
practitioners even express hope \reflist{59729159}{1072952}{34657563}{53376098}{41135403}.}{1.19}
In a post of a practitioner
looking for a way to test a WebAPI,
they contemplate that
\citedata[Back when WCF was the coolest thing, I did tests like this {[...]}.
All programatically. It worked like a charm]{25325133}~\ref{fcode:16}.
Even though they experience difficulties~\ref{fcode:13},
explaining that \citedata[for some reason {[it]} is REALLY hard to get to work
(as in, I haven't succeeded yet)]{}~\ref{fcode:13},
they do not seem to be discouraged and eventually find a
solution that works for them.
Another practitioner mentions that
\citedata[in Katalon {[there]} is a very nice way to
parameterize the selectors for GUI elements]{52539907}~\ref{fcode:16},
searching for a way~\ref{fcode:9} to make their
testing code cleaner.
Yet another practitioner judges enthusiastically~\ref{fcode:7} that
\citedata[{[validating the correctness of every component in their system is]}
obviously going to be quite a lot of work!
It could take years, but for this kind of project it's worth it]{1006189}~\ref{fcode:7},
also emphasizing that they already
\citedata[have a very comprehensive unit-test suite]{}~\ref{fcode:7} and
going so far as defining what
they believe to be meaningful tests~\ref{fcode:10}.
\revised{%
We find that a commonality of positive posts
is a sign of confidence of practitioners, or a
trust in tools or methods
that is grounded in positive experience~\ref{fcode:16}~\reflist{67709670}{46177956}{14961412}{1072952}.}{1.19}
We also identify that
ambition~\ref{fcode:3} and aspiration~\ref{cat:4}
in positive posts
is connected by practitioners
to their long term goals.
One practitioner contemplates that
\citedata[the code works `properly' {[...]} but {[they]} think automated tests would be good for the
longevity of the program]{48113464}~\ref{fcode:7}, and another reports that they are
\citedata[starting a new project, that promises to be much bigger and more
involved than anything {[they]}
have done in the past.]{6684337}~\ref{fcode:4}, which motivates them to
\citedata[keep a good workflow with {[their]} test and make sure {[they are]}
not creating gaps in {[their]} testing as {[they]} go]{}~\ref{fcode:9}.
As indicated in \Cref{fig:theory},
it is experience and knowledge
that gives those practitioners
an extra degree of trust and confidence,
from which an aspirational attitude~\ref{cat:4}
towards testing
seems to emerge.
\revised{Their attitude enables them
to reflect~\ref{cat:3} on
and explore~\ref{cat:4}
solutions for long term goals \reflist{4659714}.
They build knowledge proactively
without experiencing setbacks
that discouraged~\ref{cat:1} practitioners report \reflist{57609818}.}{1.19}
On the left side in \Cref{fig:theory} we visualize
that exploration~\ref{cat:2} and reflection~\ref{cat:3}
contribute to building knowledge
that will eventually allow them
to build trust and confidence.
But, more crucially,
seen at the top of the figure,
we indicate that it is the context
in which the ambition to test arises,
that determines the sentiment towards testing
when they engage in this process of building up knowledge.
More concretely,
\revised{%
when their environment and experience
gives them confidence and if their ambition
is grounded in an aspirational attitude,
they remain positive \reflist{1006189}{3340677}{23062243}{16938742}{53657417}{1072952}{4659714}.
But when their ambition to test
emerges in situations when the complexity
of their software projects begins to overwhelm them,
the process of
reflection~\ref{cat:3} and exploration~\ref{cat:2}
is negative \reflist{37527179}{67746901}{58840818}{4991264}{7960832}{25325133}{6475042}{18941509}.}{1.19}
Testing is then perceived as an obstacle
that might even push complexity further
and not as something that is good for the future of a project.

\begin{textbox}[Trust and Confidence - Degrees for Aspiration]{}
    Knowledge and experience in testing practices
    allow practitioners to aspire
    and enables them to consider and realize long term goals.
    It also enables them to reflect on their practice
    and explore new possibilities in
    a positive light.
    When exploration and reflection of testing practices
    are however motivated by pressure,
    for example an increase in complexity of a project,
    which rendered manual testing impossible,
    their ambition might be abandoned.
    Testing then turns into yet another obstacle.
\end{textbox}

\section{Discussion}
The qualitative analysis of 200 \stackoverflow{} posts
revealed many different facets of software testing to us.
In this section, we revisit our research questions
in the light of these observations,
their implications, and the recommendations we draw from them.
We then present threats to the validity of these findings and
close the chapter elaborating future work,
that will open the next stage of our grounded theory research.
Before revisiting our research questions
and elaborating future work,
we want to turn the focus once more
on the filtering process that yielded the dataset
that was analyzed in this paper.

\subsection{Semi-automated filtering of datasets for qualitative and quantitative research}
To narrow down our qualitative analysis
of the \stackoverflow{} dataset
we have used a semi-automated two-step process.
We first filtered the dataset using tags
and then employed sentiment analysis tools
to extract posts
which contain sentimental expressions.
\revised{%
We therefore consider the first,
tag based filtering approach
that is inspired by \citet{yang_what_2016}
suitable for qualitative studies like ours.
The low failure rate of the method in our case
suggests that the approach is also suitable
for quantitative studies of testing posts on \stackoverflow{}.}{1.18}

Regarding the second step,
for which sentiment analysis tools were used,
our evaluation is more differentiated.
Our analysis supports previous observations by
\citet{lin_sentiment_2018} and \citet{sengupta_learning_2020}:
authors on \stackoverflow{} indeed tend to discuss technology
in an objective, non-sentimental way.
Our analysis of 25 randomly selected (only tag-filtered) posts
indicates that authors who express sentiment
when asking questions about testing topics
on \stackoverflow{}
are more often expressing negative sentiment than positive.
Out of those 25 posts,
not a single one contained positive sentiment.
In the light of those observations we argue that sentiment analysis
indeed supported the goal to extract a subset of posts
that contains both positive and negative sentiment.
Deliberately extracting positive and negative sentimental posts
provided an improvement in terms of balance in sentiment.
In other words: a random selection would have only provided very few positive posts.
However, we do not consider our approach applicable for
quantitative studies
where results and implications are directly discerned
from the output of
sentiment analysis tools.
The accuracy of predictions
for sentiment
was simply not accurate enough
to provide meaningful insights
when only evaluating numbers.
Posts predicted as positive and negative
only turned out to be correct in 50\% of all cases (50 out of 100).
In 5 cases the sentiment was even the opposite of what was predicted.
We also learned that the sentiment analysis pipeline is
most accurate in identifying neutral posts.
Out of 25 samples
that were predicted to be neutral
only 2 contained sentiment.
Depending on the research question,
an approach to identify content with neutral sentiment
could therefore yield good results.
We identified that 28 posts of our dataset
were too short for meaningful analysis.
For studies similar to ours we recommend
to exclude short posts.
Posts are more likely to contain subjective
opinions and valuable content,
when they contain more than 2 paragraphs of text.
\revisedtwo{%
Our experience with analyzing the dataset by focusing on sentiment
taught us that finding the right approach and
selecting the right tools is challenging.
We acknowledge that low accuracy
of the tools we used
is also due to the choices we made.
For example,
instead of using a training dataset
containing sentences,
we could have used a dataset with paragraphs~\cite{wang_extracting_2019},
and instead of focusing on sentiment we could
have focused on emotion detection~\cite{novielli_gold_2018}.
The choices we made were founded on the literature
that was known to us at the time.
In the meantime however,
\citet{lin_opinion_2022} published a literature review
that contains a guideline
for the appropriate usage of tools
and approaches for opinion mining in software engineering.
We can only encourage using
their recommendations to navigate the field
and to gain confidence in making the right choices.}{1.2}

\subsection{How and why is sentiment expressed}
We set out with our analysis of \stackoverflow{}
posts to investigate how practitioners express sentiment
in the context of software testing
and which factors play a role when sentiment is expressed.
We identified 22 codes
which describe different expressions
that are used by practitioners
on \stackoverflow{}.
\begin{textbox}[RQ1]{}
    How do software engineers express sentiment about testing on \stackoverflow{}?\\
    \hdashrule[0.5ex]{1.01\textwidth}{.5pt}{1mm}\\
    \revised{%
    In sentimental posts on \stackoverflow{} practitioners are
    referring to external information like blogs or documentation,
    they reassure readers,
    share their ambition and subjective judgement
    of the value of testing practices and tools,
    compare different approaches,
    inquire for workarounds or new ways
    to solve a problem,
    admit their own lack of knowledge and their mistakes,
    reflect experiences,
    contemplate failure
    and sometimes exclude solutions that could solve their issues.
    Sentiment is expressed when
    desperation,
    unexpected behavior, uncertainties,
    complex issues, missing capabilities,
    or a willingness to improve is described.}{3.9}
\end{textbox}
The categorization of posts
has allowed us
to take our analysis
beyond the level of expressions.
We developed the four mayor categories
discouragement, exploration, reflection, and aspiration,
which illuminate factors that can lead to sentimentality.

\newpage
\begin{textbox}[RQ2]{}
    Which factors affect sentiment of software engineers towards testing practices?\\
    \hdashrule[0.5ex]{1.01\textwidth}{.5pt}{1mm}\\
    \revised{
    Lack of experience and knowledge,
    especially in complex environments is often indicated in posts with negative
    sentiment on \stackoverflow{}, when practitioners describe discouraging experiences.
    Trust and confidence into practice
    and understanding of long term goals
    on the other hand
    gives practitioners space for aspiration,
    expressed with positive sentiment.
    Practitioners who explore testing express negative sentiment
    when they experience unexpected behavior and positive sentiment
    when they are inspired by resources
    like books and blog entries.
    When reflecting on their practice,
    an awareness of their own blind-spots
    allows practitioners to be positive,
    while ambiguity,
    when practitioners are completely in the dark,
    is reflected negatively.}{3.9}
\end{textbox}

Going beyond this analysis
which highlights factors that lead to sentiment,
we presented a preliminary theory that
suggests how those factors go hand in hand in manifesting
sentiment around testing.
The preliminary theory also describes situational elements that seem to lead to sentiment.
\medskip
\begin{textbox}[Preliminary Interpretive Theory]{}
    \revised{%
    On \stackoverflow{} we see complexity and aspiration
    as important factors that make people ambitious about testing.
    Complexity of projects can make manual testing impossible and motivates (or forces)
    practitioners to use testing.
    Trust and confidence in testing practices
    on the other side makes people aspire to pursue
    long term goals using testing practices.
    In both cases experience and knowledge influences
    whether this ambition leads to a positive
    or negative experience.}{3.9}
\end{textbox}
\subsection{Implications}
\revised{%
The results of our analysis
of \stackoverflow{} posts about software testing
carries implications for education of software developers,
and management of software development teams.
Based on the data we have seen,
we hypothesize that
the implementation of automated testing practices
in simple projects,
when manual testing is still possible,
could allow an iterative development of testing skills
while reducing the likelihood of discouraging experiences.
Having obtained these skills,
we argue,
would then also influence
the experience of testing complex systems in a positive way.
Rejecting or approving this hypothesis could
help to clarify the role that teaching of software testing
can have in the early stages of software engineering careers (e.g., in undergraduate courses of universities).
Connected to this hypothesis,
our preliminary theory suggests
that on \stackoverflow{},
testing practices are perceived
as especially valuable
when the complexity
of a software project grows.
Refining and testing this theory in other contexts
could generate new insights into how
practitioners and students of software engineering
can be motivated to learn software testing.
\citet{pham_enablers_2014} for example identify the
same issue in a study with bachelor students.
Their study confirms that the perception of the complexity of code
affects students' motivation to practice testing.
They also report that students see the cost of
testing but fail to understand its benefit as projects are
often not critical or complex enough.
(Re-)introduction of testing practices,
when complex software development methods are taught,
so we hypothesize,
could teach students the value of software testing.
Introducing testing practices like mocking
in the context of distributed systems
and socket programming
is one example.
Regarding managers of software engineering teams,
our preliminary theory implies
that giving employees time and space to
develop simple test cases for simple projects is beneficial.
Being comfortable with simple test practices,
practitioners seem to gain confidence and trust.
As a recommendation that should be tested in future work,
we suggest that
the development process should allow a steady
increase of complexity
instead of tackling huge challenges directly.
The words of one author reflecting his work in a project
where they introduced testing echoes this last implication of our interpretation:
\citedata[While I no longer work on this project {[...]},
I think it gave me some enormous
insight into how bad some projects can be written,
and steps one developer can take to make things a lot cleaner,
readable and just flat out better with small,
incremental steps over time.]{1064403}}{3.9}


\subsection{Threats to validity}

Our systematic analysis of 200 \stackoverflow{} posts
has led to insights that have enabled us
to formulate preliminary hypotheses to answer our research questions and
an interpretive theory.
In this section we present the threats to the validity of our findings.

\subsubsection{Internal Validity}

To select samples from the \stackoverflow{} dataset
we filtered using user-assigned tags
and the sentiment analysis tools SentiCR and RoBERTa.
The dataset from \citet{lin_sentiment_2018},
which we used to train the tools,
was evaluated by \citet{zhang_sentiment_2020}, who report
macro- and micro-averaged F1-scores of $0.59$ and $0.82$ for SentiCR and
$0.80$ and $0.90$ for RoBERTa respectively.
\revisedtwo{%
However, their evaluation
was done with a dataset of sentences
and not at the level of paragraphs.
We do not know if applying the tools on paragraphs, like we did,
leads to poorer performance.}{1.2}
We combined both tools to reduce inaccuracy
as suggested by \citet{zhang_sentiment_2020}.
We only selected posts
that were classified
with the same sentiment polarity
by both tools.
We checked the accuracy of the filtering approach
by including and evaluating two groups of test samples
in our analysis (25 random and 25 neutral posts)
and classifying the sentiment of each post.
Even though the precision of the tools combined
provided only a 50\% accuracy for positive posts,
we argue that the inaccuracy does not pose a threat to our results.
The results presented in this paper were produced
by deep and thorough qualitative analysis
for which the sentiment analysis
was only a tool to narrow down the focus.
The accuracy has no direct influence on the results of our analysis.
To avoid mistakes in the implementation of the sentiment analysis tools, we used
the open-source implementation of SentiCR from the replication package of \citet{zhang_sentiment_2020}
\footnote{\href{https://github.com/soarsmu/SA4SE}{\faIcon{github} GitHub sorasmu/SA4SE}},
and the open-source library PyTorch
\footnote{\href{https://pytorch.org/hub/pytorch\_fairseq\_roberta/}{\faIcon{python} PyTorch fairseq/roberta}}
which provides an implementation of roBERTa.

To extract posts from the dataset that are relevant to software testing
we extended an existing open-source tool
\footnote{\href{https://github.com/SkobelevIgor/stackexchange-xml-converter}{\faIcon{github} GitHub SkobelvIgor/stackexchange-xml-converter}}.
With our extension of the tool we first filtered for all post
with a tag that includes the word \textit{testing}.
We then generated an include list of tags by manually removing all irrelevant tags
that occurred in this subset of posts.
Starting with a generic wild-card
and then snowballing to generate
a more accurate list of tags
was found to be a
valid method by \citet{yang_what_2016}.
Errors in the implementation of the filtering tool
and mistakes during the manual selection of tags
pose a possible threat
to the validity of our results.
To reduce the chance of implementation errors
we only made minimal changes
to the open-source software
that was used for filtering.
To minimize errors in the manual tag selection process,
the final list was reviewed by two
software engineering researchers
who were otherwise not involved in this study.

\subsubsection{Experimenter Bias}
We took measures to ensure that the influence
of the authors' subjectiveness on the results of this paper
stays within the boundaries of what is
reasonable and expected in the context of a constructivist GT study.
It is possible that the authors made mistakes in the interpretation of the dataset.
\revisedtwo{%
To reduce the likelihood of a misinterpretation
that would pose a threat to the validity of
our results, the interpretation of the data
recorded in memos and developed into sentiment classification, codes, categories and theory
was discussed between the first and second author.
Disagreements were resolved in a cooperative manner.}{3.1}
We do not provide a quantitative analysis
of this process of reliability verification
as such an analysis would suggest a level
of objectivity that we do not want to claim~\cite{mcdonald_reliability_2019}.
Aligned with our epistemological stance and the interpretive nature of constructivist GT,
we instead acknowledge our biased perspective.
\revised{Instead of claiming a high level
of absolute objectivity,
we argue that
taking the view from nowhere,
would not be appropriate
to answer the research question that we propose.
Instead, we present a transparent account
of the grounds on which our interpretation rests.
We use pertinent quotes and provide references to original documents
whenever we explain our interpretations.
The reader is invited to go through all the references in the text
and the analyzed material
that we provide with our replication package.}{3.7}
Inspection of the material should reveal to the reader that
we only make the material to speak for itself~\cite[coded-dataset.qdpx]{swillus_replication_2022}.
High involvement with the data,
enabled by following the systematic strategies of constructivist GT,
and not our preconceptions led to what we present in this paper.

We use sentiment analysis tools to filter the \stackoverflow{} dataset.
It allowed us to narrow down the dataset to what is relevant for our study.
To ensure that our own, manual evaluation of sentiments of posts and expressions
is not biased by the outcome of this tool-based classification,
documents were analyzed in random order and the results
of the tool's classification were hidden during analysis.

\subsubsection{External Validity}
Qualitative research searches for a deep understanding of the particular.
Knowledge generated from such research is context dependent.
We therefore can not claim
that the preliminary result that our analysis
produces has a high external validity
that goes beyond the scope
of the \stackoverflow{} community.
\stackoverflow{} posts, which are non-interactive documents,
cannot provide a full or \textit{thick}
description of sociological circumstances~\cite{geertz_interpretation_2017,fielding_virtual_2008}.
In other words: \stackoverflow{} posts
only provided us a shallow view
of the circumstances
that practitioners experience;
there are many things we are
unable to see through
an analysis of \stackoverflow{} posts.
By sharing our preliminary interpretive theory
we motivate inquiries that add more depth.
More in-depth inquiries that either challenge
the generalizability
of what we have learned on \stackoverflow{}, or
extend on it to fit a broader context
than the one we investigated.
To broaden the context of the posts,
we considered comments, edits, and links
that are referred to in posts and
evaluated post's edit-history
and the profiles of users that posted content.
\revised{Further,
    the conclusions that allowed us to construct
    the results of this paper
    are based on the qualitative analysis of
    a small part of the full \stackoverflow{}
    dataset.
    As analyzing the full dataset is not feasible,
    we choose to focus our analysis
    on a fraction of sentimental posts.
    By not analyzing the whole dataset we
    risk missing details that could lead
    to different interpretations and hence different theories.
    We reduced this risk by consecutively adding posts
    to our analysis until we reach a point,
    when the analysis of further posts does not
    reveal any new answers to
    the research questions we pose.
\revisedtwo{%
    We are aware that reaching such a point does not
    rule out the possibility
    that adding more posts
    can reveal new insights.
    It only signals
    that the effort required
    to obtain these insights
    gets disproportionate.
    Instead,
    concluding at this point allows moving forward to
    obtain insights from richer sources of data.}{3.5}
    Our analysis concluded in this way
    after reviewing 200 posts.}{3.5}

\revised{\subsubsection{Construct validity}
    We investigate the role of sentiment
    in software testing posts
    to learn about
    the experience of software developers
    when they practice software testing.
    We use sentimentality
    as a construct and proxy
    to analyze content
    that goes beyond
    technical discussions
    and touches on this experience.
    By analyzing sentimental \stackoverflow{} posts
    we infer interpretations
    about how sentiments come about
    and how they affect testing practices.
    The root causes
    for sentiment of practitioners
    are manifold
    and might be due to variables
    which we were not able
    to consider in our investigation.
    This poses a threat to the validity of our results.
    We reduced this threat
    by analyzing the data qualitatively,
    taking contextual information of posts
    like comments, edit history
    and the time it took
    for the question to be answered
    into account.
    We are therefore not only relying
    on sentimentality
    as a variable
    to understand
    what affects practitioners.}{3.7}


\subsection{Future Work}
The analysis described in this paper
brought us closer
to understanding what arouses sentiment in practitioners
in the context of testing.
However,
as mentioned in the threats to external validity,
the implications we present
need to be taken
with a grain of salt.
Before suggesting which steps can be taken
to raise our work to a higher level of maturity,
we reflect on the
limitations of the analysis presented in this paper.

\subsubsection{Limitations}
\label{sec:limitations}
Stack Exchange, the parent website of \stackoverflow{}, provides insights about \stackoverflow{}
by conducting an annual user survey.
Their surveys' results and independent research about diversity on the platform
reveals that the user base lacks diversity
when it comes to ethnicity and gender~\cite{ford_paradise_2016}.
In their own report it is stated that people of color are
underrepresented among professional developers on \stackoverflow{}
and that the company has considerable work to do,
to ensure the platform is inclusive\footnote{
\href{https://insights.stackoverflow.com/survey/2021\#section-demographics-gender}
{\faIcon{stack-overflow} insights.stackoverflow.com/survey/2021\#section-demographics-gender}}.
According to \citet{vadlamani_studying_2020}, and~\citet{zagalsky_how_2016}
it is not only ethnicity
and gender, but also professional factors
that are strong reasons for (a lack of) engagement in the community.
They lead to an \textit{expert-bias}
as novice contributers may even be confronted with
subtle or overt bullying on \stackoverflow{}.
Another bias is introduced through strict community guidelines\footnote{
\href{https://stackoverflow.com/help/how-to-ask}
{\faIcon{stack-overflow} stackoverflow.com/help/how-to-ask}}.
During our investigation we were
directly confronted with this limitation.
Two posts that were rich in sentiment
were closed because they violate
the community guidelines.
In one of those post,
the message posted by a moderator reads:
\citedata[as it currently stands, this question is not a good fit for our Q\&A format.
We expect answers to be supported by facts, references, or expertise, but this question
will likely solicit debate, arguments, polling, or extended discussion]{16938742}.
In the other post,
an author who has \citedata[been banging {[their]} head against the wall
trying to understand {[...]} concepts for a week]{2978843}
simply suggested
a \citedata[very understandable and simple]{} explanation
so that others can also enjoy an `aha' moment.
Examples like this make it evident that
practitioners cannot express themselves freely on \stackoverflow{}.
When they post exclusively sentimental content
or ask questions that provoke discussion,
they are sanctioned.
The aforementioned post also suggests another limitation:
practitioners posting on \stackoverflow{} are biased towards negativity.
What is discussed on \stackoverflow{} are problems.
If there is no problem to solve, the post is closed.
Success stories or exclusively positive accounts of practitioners on \stackoverflow{}
are therefore rare.

\subsubsection{Theoretic sampling}
Early stages in grounded theory
are supposed to open up discussion and
motivate for focused inquiries to follow.
Theories mature
as they are refined
and backed by
collection and analysis
of more data.
In grounded theory,
this crucial process is called
theoretic sampling~\cite{charmaz_constructing_2014}.
Apart from refining, verifying or rejecting our theory,
such a focused collection of samples can answer questions that we
derive directly from our analysis.
\begin{enumerate}
    \item
        If ambition to test arises when practitioners
        are suddenly confronted with overwhelming
        project complexity,
        how do project management frameworks like Agile
        affect adoption of testing methods compared to
        projects that use long term fixed planning?
    \item
        How are practitioners first confronted with testing practices?
        How does this first encounter with testing in a professional setting
        influence their ambitions to adopt testing in other contexts?
    \item
        If the complexity of projects under test and the required complexity
        of techniques to test them grows proportionally like our preliminary theory suggests,
        how do developers of testing tools relate to this connection in terms of
        provided documentation and design of tools?
    \item
        The analysis showed that sentiment around testing highly depends on context.
        In this study we looked at expressions of practitioners.
        How do researchers and educators in software engineering relate to testing in
        comparison to what we observed in our study?
        How does ambition differ, especially in cases where they have not been
        confronted with the factors that cause discouragement which we described in this paper?
\end{enumerate}
As we highlight in \Cref{sec:limitations},
the dataset which was analyzed in this paper
only provides a narrow perspective
on the lived experience of practitioners.
While \stackoverflow{} provides insights into what testers
do outside their Integrated Development Environment (IDE),
it only rarely provides insights into what testers do when they are
not working on their computer.
Posts rarely describe the social world in which testing is practiced.
Derived from the things we did not see in the dataset, we propose the following questions for future inquiries:
\begin{enumerate}
    \item How does the social context of individuals affect
        sentiment of testers
        when they are exploring
        or reflecting experiences?
    \item Which role
        does the experience of peers play
        in shaping the testing experience
        of individual practitioners?
    \item How do practitioners express sentiment
        about testing in informal settings?
    \item How do practitioners express sentiment online,
        when ambiguous and sentimental content which provokes
        discussion is not sanctioned but encouraged?
\end{enumerate}

In order to investigate the above questions,
we propose different approaches.
Through a \textit{quantitative analysis} of \stackoverflow{},
\citet{alshangiti_why_2019} revealed that different
challenges in the field of machine learning are present
because implementation of application requires a wide set of skills.
More concretely, they suggest that data preprocessing is especially challenging
as it is often overlooked in education of practitioners.
A quantitative content analysis like the one of \citet{alshangiti_why_2019}
about testing posts on \stackoverflow{}
could identify aspects
of testing that are difficult
to handle for practitioners on a more technical level.
Further \textit{qualitative studies}
of non-interactive documents
from platforms like Reddit\footnote{
\href{https://reddit.com/r/softwaretesting}{\faIcon{reddit} Reddit /r/softwaretesting}}
or Twitter\footnote{\href{https://twitter.com/search?q=\%23softwaretesting}{\faIcon{twitter} Twitter \#softwaretesting}},
which encourage sentimental and ambiguous content,
can complement our analysis on a non-technical level.
Conducting a \textit{meta analysis} of publications
on socio-technical aspect of software testing
is another way of grounding our work in more theoretical and empirical
data that others investigated in the past.
But most crucially,
we want to meet practitioners where
they are confronted with testing practices.
\textit{Field studies}
in which individuals
or groups of practitioners are observed
and interviewed
during practice
can provide insights
that go beyond
what non-interactive documents
can reveal.
Direct observations of practitioners will
provide crucial insights into
lived experience
that allow the formulation of a mature theory.

\section{Related Work}
With our investigation of sentimental posts on \stackoverflow{},
the categorization of posts
and the development of a preliminary theory
we highlighted different aspects
that influence motivation of practitioners,
the effect of emotions on practice,
and the role of software testing
as a part of software development.
In this section we relate our findings
to what others have uncovered
in relation to those topics.

A study by \citet{graziotin_what_2018}
emphasizes the detrimental effects
that unhappiness can have on software engineering practitioners.
Some of what they describe what happens when developers are (un)happy
is relevant to our paper.
According to their report,
developers distance themselves from tasks to
which their unhappiness relates.
Our analysis reveals that confrontation with testing can
under some circumstances cause negative feelings of discouragement.
Discouragement can thus lead to withdrawal from testing
resulting in process deviation and reduced code quality.
On the positive side,
findings of \citet{graziotin_what_2018} show that
emotions related to happiness like aspiration
increase process adherence and stimulate creativity,
leading to a stronger commitment to writing tests.

A literature review by \citet{beecham_motivation_2008}
compares the findings of 92 papers
about the topic of motivation
of software engineers from the 1980s to 2006.
The review highlights that software engineers
display a very high need for growth
and that they are concerned about learning new technology.
Software engineers are motivated
by the exploration of new techniques
and want to work on identifiable pieces of quality work.
According to the review,
problem-solving and the confrontation with challenges
can be an enhancing factor for motivation.
While those factors are present in many studies,
the literature review concludes that the
needs of software engineers
are highly dependent on the context
of individuals.
Our study confirms this conclusion.
Exploration can increase motivation or ambition
in the case of software testing,
but we indeed see that
whether challenges or exploration
lead to increased motivation
highly depends on context.
Contrary to the studies included in the review,
we see that a confrontation with challenges
can also lead to discouragement.
Our results on this aspect are more aligned
with the results of a qualitative study by
\citet{sharp_models_2009}, that suggests that challenges,
even when mentioned as a reason to stay in the job,
are not so much a factor that gives practitioners satisfaction.
Not challenges,
but creativity and being able to make a difference
is what makes software engineering worthwhile~\cite{sharp_models_2009}.
Similarly, \citet{meyer_today_2021} found out
that on \textit{good} workdays,
developers make progress
and create value
for projects they consider meaningful.
On good days,
they spend their time efficiently,
with little administrative work, and infrastructure issues;
what makes a workday typical and therefore good is primarily
assessed by the match between developers'
expectations and reality~\cite{meyer_today_2021}.
Two things here relate to our own findings.
First, we also find that practitioners who already identify testing
as good and meaningful practice,
for example because they are motivated by books or blogs about testing,
are indeed ambitious and aspirational about testing.
Second, we also see that challenges
created by infrastructure issues,
for example in complicated development environments
lead to discouragement because of unexpected behavior.
With a survey study conducted
in multiple companies \citet{runeson_survey_2006}
also found supporting evidence for the negative impact
of unexpected challenges caused by complexity.
A good integration of unit testing
into the internal tool landscape
that is provided by the company
is key for the adoption of testing.
However, this integration is especially hard
when the modules under test
interact with a complex system state
or a complex system environment~\cite{runeson_survey_2006}.
When an integration of testing into practice is
too challenging it is mostly
perceived as de-motivating for software developers.
In this context,
\citet{daka_survey_2014} report that
practitioners rank the isolation of testing code
as one of most challenging tasks.
Crucially, it is perceived
as a difficult challenge
more often by novice software developers.
We see the same in our investigation.
Our analysis suggests that inexperienced practitioners
are often discouraged from testing by complicated
environments in which an isolation
of the method under test becomes difficult.
\revised{On the other hand,
aligned with our results,
\citet{pham_enablers_2014} identified that
novice developers adjust their testing effort
according to the perceived complexity of code.
A project has to be complex to
warrant testing to be beneficial.
Complexity can thus, as we saw on \stackoverflow{} as well,
be a motivating factor.
\citet{pham_enablers_2014} and \citet{daka_survey_2014} also report
that developers' feelings about unit testing are often negative.
Concretely,
only half of the practitioners interviewed by \citet{daka_survey_2014}
had positive feelings about testing
and students interviewed by
\citet{pham_enablers_2014} were not fond of testing because to them
writing tests did not feel like an accomplishment.
Some students even developed an anxious attitude towards testing.}{3.9}
This aligns with our observation in so far that we saw
an overwhelming amount of negative posts in random samples.
A general negative bias towards testing could therefore also
be an explanation for the high amount of negative post
that we saw in our dataset.

In relation to \citet{sharp_models_2009} and \citet{meyer_today_2021}
and their finding that meaningful contributions
and being able to make a difference are important.
However, from
our own work it is not evident that testing in itself is
always recognized as a meaningful contribution to projects
by practitioners and their peers.
Positive ambitions mentioned in posts on \stackoverflow{}
mostly seem to be self-aroused for example through engagement with
inspiring resources like books or blogs.
\citet{daka_survey_2014} indeed identified that peer pressure is only rarely
mentioned as a motivating factor to write unit tests;
the driving force for a developer to use unit testing
is supposedly their own conviction.

Finally, \citet{Kasurinen_how_2011} investigated
how new testing practices
are adopted by companies
and found out that
when confronted with new techniques
that could improve testing processes,
most companies are not interested
in adoption
if there is no first-hand knowledge
in the team or company.
Only rarely they do give new practices a try,
and if they do,
they only evaluate new techniques in small projects.
However, \citet{Kasurinen_how_2011} also report
that companies adopt new techniques when clear
need arises.
According to the theory they propose in their study,
development of processes only happens
when the existing process
obviously has a need to develop;
required resources for adoption
of new practices like testing need to be justified.
Our preliminary theory has
at its core
this very point.
We observe on \stackoverflow{},
that an increase of complexity of a project
leads to spontaneous adoption of testing practices.
While it is not clear from the report of \citet{Kasurinen_how_2011},
what the motivation or rational reason of a company that
evaluates testing practices in small projects is,
a suggestion could be taken from our own study.
We suggest that evaluation of techniques
in small projects leads to an advantage
when the need for those techniques can no longer be ignored.
In other words,
first-hand knowledge should develop
in a company before it is really needed.


\section{Conclusion}

In this study we set out
to understand the sentiments of
software engineers regarding software testing
in the context of the popular
question and answer platform \stackoverflow{}.
In order to do so, we have
used a semi-automated approach to detect
sentiment in Stack Overflow posts.
In particular, we start
out by using automatic sentiment analysis tools to classify posts,
after which we perform an in-depth,
qualitative analysis.

Through this in-depth study of 200 posts
we find that developers
are in fact sentimental about software
testing on Stack Overflow;
we find that they express
their sentiment when
unexpected behavior, uncertainties, complex issues,
missing capabilities,
or a willingness to improve is part of the post.
Additionally, we have observed that
lack of experience and knowledge,
especially in complex environments
can lead to a negative sentiment. 
On the other hand,
software engineers express positive sentiment
when they have trust and confidence
in their practice,
especially if they have an understanding of long term goals of their projects.

\revised{%
Through the observations that we have made,
we construct a preliminary interpretive theory that
explains how a projects' complexity
and the tacit knowledge of individuals
shapes the experience and attitude of practitioners
in the context of software testing.
Practitioners, we argue,
get motivated to practice software testing
as the complexity of their project increases.
Reaching that point without enough knowledge
of testing practices leads to discouraging experiences.
We argue that testing practices
are also seen by practitioners
as something to aspire to,
especially when considered
for example
in the context of
long term goals.
This has implications for both
the education of software engineers,
and for managing software development
teams that engineer complex software.
Our findings suggest that taking both
motivation and complexity
into account
in future studies
of software testing practices
can reveal more about
practitioners' sentimental perspectives.
Our preliminary results show that
an investigation of the motivation
and capabilities of software engineers
to engage in effective testing practices
needs to go beyond the analysis of
technical tools and their usage.}{1.17}

We acknowledge that we need to
extend and deepen our interpretive theory,
and our overall understanding of software engineers'
sentiments towards testing.
In particular,
in our future work we envision to study the
social context and its relation to sentiment,
the connection to the experience levels of software engineers,
their sentimental expressions in informal settings,
and finally how project management culture influences
attitudes and motivation of individual software engineers in the area of testing.

\section*{Acknowledgements}
This research was partially funded by the Dutch science foundation NWO through the Vici ``TestShift'' grant (No. VI.C.182.032).

\balance
\bibliographystyle{ACM-Reference-Format}
\bibliography{references}

\end{document}